\documentclass[preprint,preprintnumbers,amsmath,amssymb,superscriptaddress]{revtex4}
\usepackage{graphicx,color}
\usepackage{amsmath,amssymb}
\usepackage{url}
\usepackage{epstopdf}
\usepackage{bbold}

\newcommand{\hs}{\hspace*{0.5cm}}

\newcommand{\be}{\begin{equation}}
\newcommand{\ee}{\end{equation}}
\newcommand{\bea}{\begin{eqnarray}}
\newcommand{\eea}{\end{eqnarray}}

\newcommand{\crn}{\nonumber \\}

\newcommand{\al}{\alpha}
\newcommand{\la}{\lambda}

\newcommand{\ga}{\gamma}

\newcommand{\fr}{\frac}
\newcommand{\bc}{\begin{center}}
\newcommand{\ec}{\end{center}}

\newcommand {\ba}{\begin{array}}
\newcommand {\ea}{\end{array}}
\newcommand{\ben}{\begin{enumerate}}
\newcommand{\een}{\end{enumerate}}

\usepackage{epsfig,graphicx}
\usepackage{bm}
\usepackage{dcolumn}
\allowdisplaybreaks
\begin{document}

\preprint{}

\title{ Decays of CP-even Higgs bosons $H \rightarrow l_il_j$ in a 3-3-1 model with neutral leptons}
\author{H.~T.~Hung \footnote{Corresponding author}}\email{hathanhhung@hpu2.edu.vn}
\affiliation{Department of Physics, Hanoi Pedagogical University 2, Phuc Yen,  Vinh Phuc 15000, Vietnam}%%%%%%%%%%%
\author{A. B. Arbuzov } \email{arbuzov@jinr.ru}
\affiliation{Bogoliubov Laboratory of Theoretical Physics, Joint Institute for Nuclear Research, Dubna, 141980 Russia}
%
%%%%%%%%%%%%%%%%%%%%%%%

\begin{abstract}
The 3-3-1 model with neutral leptons contains four CP-even Higgs bosons when some constraints are imposed on the Higgs potential. Two Higgs bosons of this type are identified with the corresponding ones in the Two-Higgs-Doublet model (2HDM), an other being more massive does not couple with similar particles contained in the Standard Model (SM). The remaining particle is assumed to have no lepton-flavor-violating couplings. The contributions at one-loop order to $\mathrm{\Gamma}(\mathrm{H}\rightarrow l_il_j)$$(H \equiv h^0_1,h^0_2,h^0_3)$ come mainly from neutral leptons leading them to depend very strongly on mass of charged Higgs boson, masses and the mixing matrix ($V^L_{ab}$) of the neutral leptons.
The numerical investigated results of $\mathrm{\Gamma}(\mathrm{H}\rightarrow l_il_j)$ in the spatial regions satisfying the experimental limit of lepton-flavor-violating decays of charged leptons (cLFV) indicate that the signal of SM-like Higgs boson ($\mathrm{\Gamma}(h^0_1\rightarrow l_il_j)$) approaches the upper bound of the experimental limit and $\mathrm{\Gamma}(\mathrm{H}\rightarrow l_il_j)$ can take large values in the case of $V^L_{ab} = U^L_{ab}(\pi/4,\pi/4,-\pi/4)$ and smaller in the case of $V^L_{ab} = U^L_{ab}(\pi/4,0,0)$. Further, some distinct properties of $\mathrm{\Gamma}(h^0_2\rightarrow l_il_j)$ and $\mathrm{\Gamma}(h^0_3\rightarrow l_il_j)$ are also discussed. 
\end{abstract}

\pacs{{\bf Last updated: \today}
 }
\maketitle
%%%%%%%%%%%%%%%%%%%
\section{Introduction}

\allowdisplaybreaks
The assumption about lepton-flavor-violating (LFV) processes are based on the precise values of mixing angles and  mass squared deviations of the active neutrinos. These data are confirmed experimentally through Refs.\cite{Patrignani:2016xqp,Zyla:2020zbs}. The two most interesting LFV processes are lepton-flavor-violating decays of charged leptons (cLFV) and lepton-flavor-violating decays of the standard model-like Higgs boson (LFVHDs).\\
Although no evidence of oscillations of charged leptons has been found, experimental limits for cLFV are continuously proposed. For example, some typical cases must be mentioned as $Br(\mu \rightarrow 3e)<10^{-12}$ in Ref.\cite{SINDRUM:1987nra}, $CR(\mu^-Ti \rightarrow e^- Ti)<6.1\times 10^{-13}$ in Ref.\cite{Lindner:2016bgg} ... But the strictest limit shown is $Br(\mu \rightarrow e\gamma)<4.2\times 10^{-13}$ as given in Refs.\cite{Patrignani:2016xqp,Zyla:2020zbs,CMS:2018ipm,ATLAS:2019erb}, which is one of the most commonly used conditions to select the parameter space region to investigate LFVHDs.\\
After the discovery of the Higgs boson in 2012 \cite{ATLAS:2012yve,CMS:2012qbp}, the signals of LFVHDs were experimentally given with increasing accuracy. The limits of these decays are given based on Refs.\cite{CMS:2018ipm,ATLAS:2019erb,ATLAS:2019pmk,ATLAS:2019old}.
%%%%%
\bea
Br(h \rightarrow \mu\tau)\leq \mathcal{O}(10^{-3}),\,
Br(h \rightarrow \tau e)\leq \mathcal{O}(10^{-3}),\,
Br(h \rightarrow \mu e)< 6.1 \times 10^{-5}. \label{hmt-limmit}
\eea
%%%%%%%%%
There are many established models that have shown the existence of regions of parameter space where the experimental limits at Eq.(\ref{hmt-limmit}) are satisfied. In particular, $Br(h_1^0\rightarrow \mu\tau)$ can reach about $\mathcal{O}(10 ^ {-4})$ in some supersymmetric \cite{Arganda:2005iq, Gomez:2017dhl} and non-supersymmetric \cite{Zhang:2015csm,Qin:2017aju} models. More interesting, by using an effective dimension-six operator and the type-I seesaw mechanism, one is possible to accommodate the CMS $h \rightarrow \mu \tau$ signal with a branching ratio of order $10^{-2}$ \cite{He:2015rqa}. However, it has also been shown that if the contributions from new particles in beyond the standard model (BSM) to $h_1^0\rightarrow \mu\tau$ decay are very small or mutually cancel then $Br(h_1^0 \rightarrow \mu\tau)$ is only about $\mathcal{O}(10 ^ {-9})$ \cite{Gomez:2017dhl,Hammad:2016bng}. Some publications also indicate the large signal of LFVHDs within the constraints of cLFV decays \cite{Herrero-Garcia:2016uab,Blankenburg:2012ex, Hung:2022kmv, Hung:2021fzb}.

Theoretically, LFVHDs can occur at tree or loop order depending on the specific interactions of each model. In BSM, the loops are more diverse due to the appearance of new particles. However, we can always distinguish into two main types: the contribution of bosons and the contribution of fermions. On the contribution of bosons, we pay attention to both the charged Higgs bosons and the charged gauge bosons. Because the contribution of the $W$-boson is strongly suppressed by GIM mechanism then the main contribution of the gauge bosons comes from the new charged bosons, which are outside the standard model. However, the masses of these particles are always constrained by $m_{Z'}$ as shown in Ref.\cite{ATLAS:2019erb}. Remaining, the contributions of charged Higgs bosons are varied and depend heavily on the new energy scales. More interesting thing is contributions of fermions, which include ordinary charged leptons, exotic leptons and neutrinos, but ordinary charged leptons are assumed to be unmixed so its contribution can be determined pretty simply. The complex part belongs to neutrinos and exotic leptons with different mixing mechanisms. To solve the masses and oscillations of active neutrinos, we can use seesaw mechanisms Refs.~\cite{Gomez:2017dhl,CarcamoHernandez:2019pmy,Catano:2012kw,Hernandez:2014lpa,Dias:2012xp,Nguyen:2018rlb,Hue:2021zyw} or otherwise as shown in Refs.~\cite{Hue:2017lak,Thuc:2016qva}. Besides, to give the mass basis of exotic leptons, we have different assumptions for large LFV effects as given in  Refs.~\cite{Hue:2017lak,Thuc:2016qva,Hong:2020qxc}. 
%%%%%%%%

Recently, the 3-3-1 models with various sources of lepton flavor violating couplings are used to investigate LFV decays \cite{PhysRevD.22.738,Chang:2006aa, Okada:2016whh,  Dias:2006ns, Diaz:2004fs, Diaz:2003dk, Fonseca:2016xsy, Buras:2012dp, Buras:2014yna, Hue:2017lak, Nguyen:2018rlb,Hue:2021zyw,Hue:2021xap,Hong:2020qxc}. 
However, these models can only provide very small LFV signals or cLFV and LFVHDs can achieve relatively large signals but in different regions of the parameter space \cite{Hue:2015fbb, Thuc:2016qva,Boucenna:2015zwa,Hernandez:2013hea}. By combining constraints to indicate suitable parameter space regions, some works have shown large signals of LFVHDs within the experimental limits of cLFV \cite{Hung:2021fzb,Hung:2022kmv}. Besides the LFV decays, the 3-3-1 models can also give large signals of other SM-like Higgs boson decays such as, $h^0_1\rightarrow \gamma \gamma$ and $h^0_1\rightarrow Z \gamma$ \cite{Hung:2019jue} .

In this work, we consider a 3-3-1 model with neutral leptons and use some constraints on the Higgs potential to give three CP-even Higgs boson. The analytic formulas of $\mathrm{\Gamma}(\mathrm{H}\rightarrow l_il_j)$ at the one-loop order are established. We use regions of the parameter space that satisfy the experimental limits of cLFV and have large signals of both $h^0_1\rightarrow \mu \tau$ decay \cite{Hung:2022kmv} and $h^0_1\rightarrow Z \gamma$ decay \cite{Hung:2019jue} to investigate $\mathrm{\Gamma}(\mathrm{H}\rightarrow l_il_j)$. We expect to find the characteristic properties of $\mathrm{\Gamma}(h^0_2\rightarrow l_il_j)$ and $\mathrm{\Gamma}(h^0_3\rightarrow l_il_j)$ in the region of space where $\mathrm{\Gamma}(h^0_1\rightarrow l_il_j)$ approaches the upper bound of the experimental limit.

The paper is organized as follows. In the next section, we review the model and give masses spectrum of Higgs and gauge bosons. We then show all LFV couplings in Section \ref{Couplings}. We calculate the Feynman rules and analytic formulas for $\mathrm{\Gamma}(\mathrm{H}\rightarrow l_il_j)$  in Section \ref{Analytic}. Numerical results are discussed in Section.\ref{numerical_results}. Conclusions are in Section \ref{conclusion}. Finally, we provide Appendix \ref{appen_PV}, \ref{appen_loops1}, \ref{appen_loops2} to calculate and exclude divergence in the amplitude of $\mathrm{\Gamma}(\mathrm{H}\rightarrow l_il_j)$.

\section{\label{331LHN} In brief of 3-3-1 model with neutral leptons}

\subsection{ Particle content}
We consider a particular 3-3-1 model as a class of general 3-3-1 models (331$\beta$) with $\beta=-\frac{1}{\sqrt{3}}$. This model is based on the gauge symmetry group $SU(3)_C\otimes SU(3)_L\otimes U(1)_X$ and the third component at the bottom of lepton triplets are neutral leptons. The parameter $\beta$ is a basis for defining the form of electric charge operator in this model: $Q=T_3+\beta T_8+X$, where $T_{3,8}$ are diagonal $SU(3)_L$ generators. In order to obtain the LFV source for strong enough influence to study LFVHD, we build a model with the same characteristics as the 331NL model mentioned in Refs.~\cite{Hue:2015fbb,Hung:2022kmv}.

There are two common ways to arrange particles into triplets or anti-triplets of $SU(3)_L$ group. Firstly, the left-handed leptons are placed in the anti-triplets and the left-handed quarks are in the two triplets and one anti-triplet, secondly the left-handed leptons are accommodated in the triplets and the two quark generations are in anti-triplets, the remaining generation settled in triplet. These two ways are shown to be equivalent when we permutate the vacuum expectation values of the scalars which responsible generate masses for gauge bosons $W$ and $Z$. This process only alters the mixing of the neutral gauge bonsons $Z$ and $Z^\prime $ \cite{Hue:2018dqf} without affecting the expected results presented in this paper. Therefore, we randomly choose the second arrangement. That way, the lepton part in the 331NL model is represented as follows: 

\begin{eqnarray}
	L_{aL}^\prime = \left (
	\begin{array}{c}
		\nu^\prime_{a} \\
		l^\prime_{a} \\
		N^\prime_{a}
	\end{array}
	\right )_L\sim(1\,,\,3\,,\,-1/3)\,,\,\,\,l^\prime_{aR}\,\sim(1,1,-1)\,,\,\,\,N^\prime_{aR}\,\sim(1,1,0),
	\label{L}
\end{eqnarray}
where $a=1,2,3$ is generation indices, the numbers in the parentheses are  the respective representations of the $SU(3)_C$, $SU(3)_L$  and $U(1)_X$ gauge groups. We use primes to represent states in the initial basis. Thus, they are distinguished from the physical states that we often refer to in later calculations. The right-handed components of  the exotic neutral leptons and the charged leptons are $N^\prime_{aR}$ and $l^\prime_{aR}$, respectively. $N^\prime_{aL,R}$ are considered as the new degrees of freedom in the 331NL model.

The quarks in the 331NL model include left-handed and right-handed particles, in which left-handed particles are classified as antitriplets or triplets, right-handed particles are accommodated in singlets of the  $SU(3)_L$ group. These arrangements are intended for anomaly cancellation \cite{Diaz:2004fs} and details are given as follows:
\begin{eqnarray}
	&&Q^\prime_{iL} = \left (
	\begin{array}{c}
		d^{\prime}_{i} \\
		-u^{\prime}_{i} \\
		D^{\prime}_{i}
	\end{array}
	\right )_L\sim(3\,,\,\bar{3}\,,\,0)\,, \nonumber \\
	&&
	d^{\prime}_{iR}\,\sim(3,1,-1/3),\,\,\,
	\,\,u^{\prime}_{iR}\,\sim(3,1,2/3)\,,\,\,\,\, D^{\prime}_{iR}\,\sim(3,1,-1/3),\nonumber \\
	&&Q^\prime_{3L} = \left (
	\begin{array}{c}
		u^{\prime}_{3} \\
		d^{\prime}_{3} \\
		U^{\prime}_{3}
	\end{array}
	\right )_L\sim(3\,,\,3\,,\,1/3)\,, \nonumber \\
	&&
	u^{\prime}_{3R}\,\sim(3,1,2/3),
	\,\,d^{\prime}_{3R}\,\sim(3,1,-1/3)\,,\,U^{\prime}_{3R}\,\sim(3,1,2/3),
	\label{quarks} 
\end{eqnarray}
where $i=1,2$ is used to represent the first two generations. The exotic quarks are $U^{\prime}_{3L,R}$ and $D^{\prime}_{iL,R}$, they have the same electrical charge as ordinary up-quarks and down-quarks, respectively.  

The scalar sector is introduced three triplets that are guaranteed to generate the masses of the SM fermions, 
\begin{eqnarray}
	\eta = \left (
	\begin{array}{c}
		\eta^0 \\
		\eta^- \\
		\eta^{\prime 0}
	\end{array}
	\right )\sim (1\,,\,3\,,\,-1/3),\,\rho = \left (
	\begin{array}{c}
		\rho^+ \\
		\rho^0 \\
		\rho^{\prime +}
	\end{array}
	\right )\sim (1\,,\,3\,,\,2/3),\,
	\chi = \left (
	\begin{array}{c}
		\chi^{\prime 0} \\
		\chi^{-} \\
		\chi^0
	\end{array}
	\right )\sim (1\,,\,3\,,\,-1/3)\,, 
	\label{scalarcont} 
\end{eqnarray}
with $\eta$ and $\chi$ have the same quantum numbers only $\rho$ has a difference. This choice is considered the simplest and still ensures the two requirements of both chiral anomalies cancellation \cite{Diaz:2004fs} and mass generation for most fermions at tree level \cite{Long:1995ctv}.

The vacuum expectation values (VEVs) of the scalar fields are chosen to ensure mass generation for the particles and reduce the free parameter of the 331NL model. Therefore, we are interested in neutral scalar components. The 331NL model exits two global symmetries, namely $L$ and $\mathcal{L}$ are the normal and new lepton numbers, respectively \cite{Chang:2006aa,Tully:2000kk}. They are related to each other by $L=\frac{4}{\sqrt{3}}T_8+\mathcal{L}$ with $T_8=\frac{1}{2\sqrt{3}}\mathrm{diag}(1,1,-2)$. So, $L$ and $\mathcal{L}$ of the neutral scalar components are given as Tab.\ref{zero_vev},

\begin{table}[h]
	%\scalebox{0.92}
	\begin{tabular}{|c|ccccc|}
		\hline
      Neutral scalars & $\eta^0$ & $\eta^{\prime 0}$ & $\rho^0$&  $\chi^{\prime 0}$ & $\chi^0$\\
		\hline
		$\mathcal{L}$ & $-\frac{2}{3}$ & $-\frac{2}{3}$ & $-\frac{2}{3}$&  $\frac{4}{3}$ & $\frac{4}{3}$\\
		\hline	
		$L$ & $0$ & $-2$ & $0$&  $2$ & $0$\\
		\hline
	\end{tabular}
	\caption{The lepton numbers of the neutral scalar components.} \label{zero_vev}
\end{table}

As a result, the normal lepton number $L$ of $\eta^0$, $\rho^0$ and $\chi^0$ are zeros. In contrast, $\eta'^0$ and $\chi'^0$ are bilepton with $L=\mp 2$. This is the difference in lepton numbers of the components in the $\eta$ and $\chi$ triplets which will lead to unwanted lepton number violation interactions relate to a heavily neutral Higgs boson ($h^0_4$)-will be mentioned later. However, we can avoid this by choosing extremely small coefficients, the result is that $h^0_4$ has almost no couplings of lepton number violation. To generate masses for exotic quarks ($U_3, D_i$), we need to break the gauge group $SU(3)_L\otimes U(1)_X \rightarrow SU(2)_L\otimes U(1)_Y $, which requires $\left\langle \chi^0\right\rangle $ to be non-zero and on the scale of  exotic quark mass. Therefore, for simplicity and without affecting this process, one choose $\left\langle \eta^{\prime 0}\right\rangle =0 $ \cite{Chang:2006aa,Long:1995ctv,Hung:2022kmv}. The next stage is to break symmetry of the electroweak group of the SM for purpose of generating mass for the SM particles, we need the VEVs of two scalar triplets ($\eta, \rho$). The number of scalar triplets needed for this is two, because fermion triplets have two different transformations. Specifically, $\left\langle \rho^0\right\rangle$ will give mass to three charged leptons, one down-type and two up-type quarks while $\left\langle \eta^0\right\rangle$ will take care of mass generation for the remaining quarks \cite{Long:1995ctv,Mizukoshi:2010ky} and active neutrinos through the effective interaction \cite{Hung:2022kmv}. Therefore, to reduce the free parameter in the model, one chooses $\left\langle \eta^{\prime 0}\right\rangle=0$. 
  
From the above comments, all VEVs in this model are introduced as follow, 
\bea \eta^{\prime 0}&=& \frac{S'_1+i A'_1}{\sqrt{2}},\hs \chi^{\prime 0}= \frac{S'_3+i A'_3}{\sqrt{2}} \crn
\eta^0=\frac{1}{\sqrt{2}}\left(v_1+S_1+iA_1\right),\;
\rho^0 &=&\frac{1}{\sqrt{2}}\left(v_2+S_2+iA_2\right),\;  \chi^0=\frac{1}{\sqrt{2}}\left(v_3+S_3+iA_3\right). \label{vevs}\eea
Selected VEVs such as Eq.(\ref{vevs}) will perform the breaking electroweak group of the model in two stages as follows:
\begin{equation*}
	{SU(3)_{L}\otimes U(1)_{X}\xrightarrow{\langle \chi \rangle}}{%
		SU(2)_{L}\otimes U(1)_{Y}}{\xrightarrow{\langle \eta \rangle,\langle \rho
			\rangle}}{U(1)_{Q}},
\end{equation*}
where they satisfy the hierarchy ${v_3 \gg v_1,v_2}$  as done
in Refs. \cite{Chang:2006aa,Dong:2010gk}. 

The most general scalar potential was constructed based on Refs.\cite{Chang:2006aa,Hue:2021xap} has the form,
\begin{eqnarray} V(\eta,\rho,\chi)&=&\mu_1^2\eta^2
	+\mu_2^2\rho^2+\mu_3^2 \chi^2 +\lambda_1\eta^4
	+\lambda_2\rho^4+\lambda_3\chi^4  \nonumber \\
	&&+\lambda_{12}
	(\eta^{\dagger}\eta)(\rho^{\dagger}\rho)+\lambda_{13}(\eta^{\dagger}\eta)(\chi^{\dagger}\chi)
	+\lambda_{23}(\rho^{\dagger}\rho)(\chi^{\dagger}\chi) \nonumber \\
	&&+\tilde{\lambda}_{12}
	(\eta^{\dagger}\rho)(\rho^{\dagger}\eta)+\tilde{\lambda}_{13}(\eta^{\dagger}\chi)(\chi^{\dagger}\eta)
	+\tilde{\lambda}_{23}(\rho^{\dagger}\chi)(\chi^{\dagger}\rho) \nonumber \\
	&&+\sqrt{2}fv_3\left( \epsilon^{ijk}\eta_i \rho_j \chi_k +\mbox{H.c}\right),
	\label{potential}
\end{eqnarray}
where $f$ is a dimensionless coefficient included for later convenience. Compared to the general form in Ref.\cite{Chang:2006aa}, small terms in the Higgs potential in Eq.~(\ref{potential}) that violating the lepton number have been ignored. This helps us to reduce the free parameters in the model derived from the self-coupling constants of the scalar fields without much influence on the later investigation of LFV processes.
\subsection{\label{Higgs_spectral} Higgs and Gauge bosons}

In the 331NL model, there are four charged scalar components including: two components in  $\rho$ triplet and each of the remaining triplet contributing one more component. Among of these, two are absorbed by the Goldstone fields ($G^\pm_W,\,G^\pm_V$) to give masses to the charged gauge bosons $W$ and $V$, the other two give rise to two charged Higgs bosons. The masses and physical state of the charged Higgs bosons given according to Refs.\cite{Hung:2022kmv,Hue:2015fbb} are :
\bea 	m_{{H_1}^{\pm }}^2 = \left(v_1^2+v_2^2\right) \left(\frac{\tilde{\lambda }_{12}}{2}+\frac{f
	v_{3}^2}{v_1v_2}\right);\hs m_{{H_2}^{\pm }}^2 = \left(v_2^2+v_3^2\right) \left(\frac{\tilde{\lambda
	}_{23}}{2}+\frac{fv_1}{v_{2}}\right), \label{massedcH}
\eea
and
	\bea \left(
\begin{array}{c}
	\rho^{\pm} \\
	\eta^{\pm} \\
\end{array}
\right)= \left(
\begin{array}{cc}
	-s_{12} & c_{12} \\
	c_{12}& s_{12} \\
\end{array}
\right) \left(
\begin{array}{c}
	G^{\pm}_W \\
	H_1^{\pm} \\
\end{array}
\right), \hs  \left(
\begin{array}{c}
	\rho'^{\pm} \\
	\chi^{\pm} \\
\end{array}
\right)=\left(
\begin{array}{cc}
	-s_{23} & c_{23} \\
	c_{23}& s_{23} \\
\end{array}
\right) \left(
\begin{array}{c}
	G^{\pm}_V \\
	H_2^{\pm} \\
\end{array}
\right),
\label{sHiggse}\eea
where $s_{ij}\equiv\sin\beta_{ij}$,~ $c_{ij}\equiv\cos\beta_{ij}$, and $t_{ij}\equiv\tan\beta_{ij}=\frac{v_i}{v_j}$.

Next, we consider to CP-even neutral Higgs bosons. Based on Eq.(\ref{vevs}), we have five real scalars initially, namely $S_1, S_2, S_3, S_1'. S_3'$. In the final state, similar as Refs.~\cite{Hue:2015mna,Hue:2015fbb}, we get four massive neutral Higgs bosons and a Goldstone boson ($G_{U}$) corresponding to the non-Hermitian gauge boson $U$. We are first consider to the mixing of $S_1'$ and $S_3'$ which creates a neutral Higgs boson $h^0_4$. Relationship in the original basis ($S'_1, S'_3$) and mass of $h^0_4$ are:
\bea   \left(
\begin{array}{c}
	S'_{1} \\
	S'_{3} \\
\end{array}
\right)=\left(
\begin{array}{cc}
	-s_{13} & c_{13} \\
	c_{13}& s_{13} \\
\end{array}
\right) \left(
\begin{array}{c}
	G_U \\
	h_4^{0} \\
\end{array}
\right),\, m^2_{h_4^0}= \left(v_1^2+v_3^2\right) \left(\frac{\tilde{\lambda }_{13}}{2}+\frac{fv_2}{ v_{1}}\right)
\label{nrHigg1}\eea
However, we have assumed no mixing between active neutrinos and neutral leptons in the 331NL model, which will be moderated by Lagrangian later, resulting in $h^0_4$ no having LFV couplings. Therefore, in the context of LFV processes, we do not consider the case of this neutral Higgs boson.

The remainders are three CP-even Higgs bosons, we will denote them as $h^0_1$, $h^0_2$, $h^0_3$. To give their mass and physical state, we apply the aligned limit, first introduced by Ref.\cite{Okada:2016whh}.
\begin{equation}\label{eq_alignH0}
	f=\lambda_{13}t_{12} =\frac{\lambda_{23}}{t_{12}}.
\end{equation}	
 Accordingly, the mass and physical state of a heavily neutral Higgs boson ($h^0_3$) are:
\bea
m^2_{h^0_3}= \lambda_{13}v_1^2 +2\lambda_3 v_3^2,\,h^0_3\equiv S_3 ,
\label{massenH}
\eea
Using the technique mentioned in Refs.~\cite{Okada:2016whh,Hung:2022kmv,Hue:2015fbb}, we give the masses and physical states of the remaining two neutral Higgs bosons as follows:
%%%%%%%
\begin{align}
	m^2_{h^0_1}&= M^2_{22}\cos^2\delta +M^2_{11}\sin^2\delta - M^2_{12}\sin2\delta,\crn
	m^2_{h^0_2}&=  M^2_{22}\sin^2\delta +M^2_{11}\cos^2\delta + M^2_{12}\sin2\delta, \crn
	\tan2\delta&= \frac{2M^2_{12}}{M^2_{22} -M^2_{11}}. \label{ang_mixingh0}
\end{align}
%%%%%
The components ($M_{\mathrm{ij}}$) depend on the parameters of the model and have the specific form:
%%%%%
\begin{align}
	M^2_{11}&= 2s^2_{12}c^2_{12}\left[ \lambda_1 +\lambda_2 -\lambda_{12}\right] v^2 +\frac{\lambda_{13}v^2_3}{c^2_{12}},\crn
	M^2_{12}&= \left[\lambda_2 c^2_{12} - \lambda_1 s^2_{12} -\lambda_{12} (c^2_{12} -s^2_{12})\right]s_{12}c_{12} v^2 =\mathcal{O}(v^2),\crn
	M^2_{22}&= 2\left( s_{12}^4 \lambda_{1} +c_{12}^4 \lambda_{2} +s_{12}^2 c_{12}^2 \lambda_{12}\right) v^2=\mathcal{O}(v^2),~v^2=v_1^2+v_2^2,  \label{eq_mixingh0} 
\end{align}
and
\begin{align}
	\left(
	\begin{array}{c}
		S_1 \\
		S_2\\
	\end{array}
	\right)&= R^T(\alpha)\left(
	\begin{array}{c}
		h^0_1 \\
		h^0_2 \\
	\end{array}
	\right)~ \mathrm{where} ~ \alpha\equiv \beta_{12}  -\frac{\pi}{2} + \delta~ \mathrm{and} ~R(\alpha) =\left(
	\begin{array}{cc}
		c_{\al} & -s_{\al} \\
		s_{\al}& c_{\al} \\
	\end{array}
	\right).\label{mixingh0}
\end{align}
%%%%%
The lightest CP-even Higgs boson, is denoted as $h^0_1$, is SM-like Higgs boson found at LHC. From Eqs.(\ref{ang_mixingh0},\ref{eq_mixingh0}), we can see that  $\tan2\delta= \frac{2M^2_{12}}{M^2_{22} -M^2_{11}}=\mathcal{O}(\frac{v^2}{v_3^2}) \simeq0$ when $v^2\ll v_3^2$. In this limit, $m^2_{h^0_1}= M^2_{22} + v^2\times \mathcal{O}(\frac{v^2}{v_3^2})\sim  M^2_{22}$ while $m^2_{h^0_2}= M^2_{11} + v^2 \times  \mathcal{O}(\frac{v^2}{v_3^2}) \simeq M^2_{11}$. In the next part, we will see more explicitly that the couplings of $h_1^0$ are the same as those given in the SM in the limit $\delta\rightarrow 0$. 

Using the invariance trace of the squared mass matrices which is obtained after rotating the mass squared mixing matrix in the original basis by angle $\alpha$, we have
\bea
2 \lambda _1 v_1^2+\lambda _{13}v_3^2+2\lambda _2 v_2^2+\lambda _{13}v_3^2t_{12}^2=m^2_{h^0_1}+m^2_{h^0_2}. \label{trace_m2h}
\eea
the $\lambda_{13}$ can be written as 
\begin{equation}\label{eq_la13} 
	\lambda_{13}=\frac{c^2_{12}}{v_3^2} \left[  m^2_{h^0_1} +m^2_{h^0_2} -\frac{8m_W^2}{g^2}\left( \lambda_1s_{12}^2 +\lambda_2c_{12}^2  \right)\right] . 
\end{equation}
Paying attention to the aligned limit, we get the parameterized expression of $\la_{23}$ as	
\begin{equation}\label{eq_la23} 
	\lambda_{23}=\frac{s^2_{12}}{v_3^2} \left[  m^2_{h^0_1} +m^2_{h^0_2} -\frac{8m_W^2}{g^2}\left( \lambda_1s_{12}^2 +\lambda_2c_{12}^2  \right)\right] . 
\end{equation}
We can choose $m_{h^0_1}$, $\delta$ - angle as input parameters, based on Eqs.(\ref{ang_mixingh0},\ref{eq_mixingh0},\ref{trace_m2h}), the dependent parameters are given as follows:	%
\begin{align}
	\label{eq_la122}
	\lambda_{12}&=-2\lambda_1t^2_{12}  + \frac{\left( s_{2\delta} +2 t_{12} c^2_{\delta} \right) g^2m^2_{h^0_1} +\left(  2t_{12}s^2_{\delta}-s_{2\delta} \right)g^2m^2_{h^0_2}}{8 s_{12}c_{12} m_W^2},\crn
	m^2_{h_2^0}&=\frac{8\left( \lambda_2-t^4_{12}\lambda_1\right)m^2_Wc_{12}^2+g^2 m^2_{h^0_1}\left[ c^2_{\delta} (t^2_{12}-1) +t_{12} s_{2\delta}\right]}{g^2\left[s^2_{\delta}(1-t^2_{12}) +s_{2\delta} t_{12}\right]} ,  
\end{align}
The other self-coupling constants of scalars will be discussed below. They should satisfy all constraints discussed in the literature to guarantee the pertubative limits, the vacuum stability of the Higgs potential~\cite{Sanchez-Vega:2018qje}, and the positive squared masses of all Higgs bosons.
%%%%%%%%%%%%%%%
Next, we are interested in gauge bosons but only for the charged part, because the neutral part has little to do with the LFV processes of the CP-even Higgs bosons. They are started from covariant derivative of the $SU(3)_L\otimes U(1)_X$ group,
\be D_{\mu}\equiv \partial_{\mu}-i g_3 W^a_{\mu}T^a-g_1 T^9X X_{\mu}. \label{code} \ee
where  $T^a$ ($a=1,8$) and $T^9$ are generators of $SU(3)_L$ and $U(1)_X$ group, respectively. They are represented by Gell-Mann matrices ($\lambda_a$) or unit matrix ($\mathbb{1}$). Detail, $T^9=\frac{1}{\sqrt{6}}\mathbb{1}$ and $T^a=\frac{1}{2}\lambda_a,-\frac{1}{2}\lambda_a^T$  or $0$ depending on the triplet, antitriplet or singlet representation of the $SU(3)_L$ that  $T^a$ acts on. We also have denoted $W_\mu^+ = \frac{1}{\sqrt{2}}(W_\mu^1-iW_\mu^2)$, as usual, $V_\mu^-=\frac{1}{\sqrt{2}}(W_\mu^6-iW_\mu^7)$ and $U_\mu^0=\frac{1}{\sqrt{2}}(W_\mu^4-iW_\mu^5)$, charged gauge bosons part is: 
\bea W^a_{\mu} T^a= \frac{1}{\sqrt{2}} \left(
\begin{array}{ccc}
	0 & W^{+}_{\mu} & U^0_{\mu} \\
	W^{-}_{\mu} & 0 & V^{-}_{\mu} \\
	U^{0*}_{\mu} & V^{+}_{\mu} & 0 \\
\end{array}
\right).
\label{gaugeboson1} \eea
We can derive the masses of these gauge bosons from the kinetic terms of the scalar fields. Their specific form are:
\be  m_W^2=\frac{g^2}{4}\left(v^2_1+ v^2_2\right),\hs m^2_{V} =\frac{g^2}{4}\left(v^2_2+v^2_3\right),\hs m^2_{U}=\frac{g^2}{4}\left(v^2_1+ v^2_3\right), \label{gmass}\ee
where we used the relation  $v_1^2+v_2^2=v^2\equiv 246^2 \mathrm{GeV^2}$ so that the mass of the W-boson in the 331NL model matches the corresponding one in the SM. 
%%%%%%%%%
\section{Couplings for LFV decays}
\label{Couplings}

 Different from the 3-3-1 models which have right-handed neutrinos, generating mass for the leptons instead of just a scalar triplet ($\rho$) \cite{Chang:2006aa,Long:1995ctv}, we have to use all triplets ($\eta,\rho,\chi$) in the 331NL model. In particular, to generate mass for active neutrinos we need to introduce an effective interaction as shown in Ref.~\cite{Mizukoshi:2010ky}. Yukawa lagrangian of lepton part is:
\be -\mathcal{L}^{Y}_{\mathrm{lepton}} = h^{l}_{ab}\overline{L^\prime_{a}}\rho l'_{bR}+ h^{N}_{ab}\overline{L^\prime_{a}}\chi N'_{bR}+ \frac{h^{\nu}_{ab}}{\Lambda} \left(\overline{(L^\prime_{a})^c}\eta^*\right)\left(\eta^{\dagger}L^\prime_{b}\right) + \mathrm{h.c.}, \label{Ylepton1}\ee
where the notation $(L^\prime)^c_a=( (\nu'_{aL})^c,\;(l'_{aL})^c,\;(N'_{aL})^c\;)^T\equiv( \nu'^c_{aR},\;l'^c_{aR},\;N'^c_{aR}\;)^T$  implies that $\psi^c_R\equiv P_R \psi^c= (\psi_L)^c$ with $\psi$ and $\psi^c \equiv C\overline{\psi}^T$ being the Dirac spinor and its charge conjugation, respectively. Remind that $P_{R,L}\equiv \frac{1\pm\gamma_5}{2}$ are the right- and left-chiral operators, we have $\psi_L= P_L \psi, \; \psi_R=P_R\psi$. The $\Lambda$ is some high energy scale.

The choice like Eq.(\ref{Ylepton1}) leads to a very natural hierarchy, charged leptons are generated masses by $\left\langle \rho^0\right\rangle$, heavy neutral leptons are produced masses by $\left\langle \chi^0\right\rangle$ and active neutrinos are taken by $\left\langle \eta^0\right\rangle$. However, we are faced with some interactions that violate the lepton number $L$ (the value corresponding to each of the fields indicated in Ref.\cite{Hung:2022kmv}), such as: $h^{N}_{ab}\overline{\nu^\prime_{a}}\chi^{\prime 0} N'_{bR}$, $\frac{h^{\nu}_{ab}}{\Lambda} \overline{(\nu^\prime_{a})^c}\left(\eta^{\prime 0}\right)^2 \nu^\prime_{b}$, ... Therefore, to avoid undesirable effects from this, we will impose very small coefficients associated with such interactions. Based on these comments, we will show some limitations in couplings of $h^0_4$. Firstly, the third term of Eq.(\ref{Ylepton1}) is used to generate masses for active neutrinos so $\frac{h^{\nu}_{ab}}{\Lambda}$ is very small, furthermore the term $\frac{h^{\nu}_{ab}}{\Lambda}(\eta '^0)^2\overline{N'_{aL}}  N'_{bR}$ violates lepton number $L$ ($L(\eta'^0)=-2$) leading to the coupling between $h^0_4$ with $N'_a$ can be omitted. Secondly, lepton number violation also occurs with term $h^{N}_{ab}\chi^{'0}\overline{\nu^\prime_{a}} N'_{bR}$ ($L(\chi^{'0})=2$), so the couplings of the form $h^{0}_4\overline{\nu^\prime_{a}} N'_{bR}$ must be suppressed. This means that the active neutrinos do not couple with neutral leptons in the 331NL model. Therefore, we are only interested in the LFV process of three CP-even Higgs bosons denoted as $h^0_1$, $h^0_2$, $h^0_3$.

The corresponding mass terms are
\be -\mathcal{L}^{mass}_{\mathrm{lepton}} = \left[ \frac{h^{l}_{ab}v_2}{\sqrt{2}}\overline{l'_{aL}} l'_{bR}+\frac{ h^{N}_{ab}v_3}{\sqrt{2}}\overline{N'_{aL}}  N'_{bR}+ \mathrm{h.c.} \right]+ \frac{h^{\nu}_{ab}v^2_1}{2 \Lambda}\left[ (\overline{\nu'^c_{aR}} \nu'_{bL})+ \mathrm{h.c.}\right]. \label{mterm}\ee
Since there are no right-handed components, the mass matrix of active neutrinos is  $(M_{\nu})_{ab} \equiv  \frac{h^{\nu}_{ab}v^2_1}{ \Lambda}$ and proved to be symmetric based on Ref.~\cite{Mohapatra:1991ng}. The mass eigenstates can be found by a mixing matrix $U$ that satisfies $ U^{\dagger} M_{\nu}U=\mathrm{diagonal}(m_{\nu_1},\;m_{\nu_2}, \;m_{\nu_3})$, where $m_{\nu_i}$ ($i=1,2,3$) are mass eigenvalues of the active neutrinos.

Experiments have not yet found the oscillation of charged leptons. The proof is the upper bounds of recent experiments for the LFV processes in the normal charged leptons are very suppressed. This is confirmed in Refs.\cite{BaBar:2009hkt,Hayasaka:2010np,MEG:2011naj}. So, it imply that the two flavor and mass bases  of charged leptons should be the same. We now define transformations between the flavor basis  $\{l'_{aL,R},~\nu'_{aL},~N'_{aL,R}\}$ and the mass basis $\{l_{aL,R},~\nu_{aL},~N_{aL,R}\}$:
\be
l'_{aL}= l_{aL},  ~~l'_{aR}=l_{aR}, \;
\nu'_{aL}=U_{ab}\nu_{bL},\; N'_{aL}=V^L_{ab}N_{bL},\quad N'_{aR}=V^R_{ab}N_{bR},
\label{lepmixing}\ee
where $V^L_{ab},~U^L_{ab}$ and  $V^R_{ab}$ are transformations between flavor and mass bases of  leptons. Need to remember, primed fields and unprimed fields denote the flavor basis and the mass eigenstates, respectively. We also use the transformation $\nu'^c_{aR}=(\nu'_{aL})^c=U_{ab}\nu^c_{aR}$. The four-spinors representing the active neutrinos are $\nu^c_{a}=\nu_{a}\equiv (\nu_{aL},\; \nu^c_{aR})^T$, resulting the following equalities: $\nu_{aL}=P_L\nu^c_a=P_L\nu_a$ and $\nu^c_{aR}=P_R\nu^c_a=P_R\nu_a$. 

The relations between the mass matrices of leptons in  two flavor and mass bases are
\bea m_{l_a}&=&\frac{v_2}{\sqrt{2}}h^l_{a},\hs h^l_{ab}=h^l_a\delta_{ab},\hs a,b=1,2,3,
\crn \frac{v_1^2}{\Lambda} U^{\dagger}H^{\nu} U&=& \mathrm{Diagonal}(m_{\nu_1},~m_{\nu_2},~m_{\nu_3}),\crn
\frac{v_3}{\sqrt{2}} V^{L\dagger}H^N V^R&=& \mathrm{Diagonal}(m_{N_1},~m_{N_2},~m_{N_3}),\label{cema1} \eea
where $H^{\nu}$ and $H^N$ are Yukawa matrices defined as $(H^{\nu})_{ab}=h^{\nu}_{ab}$ and $(H^{N})_{ab}=h^{N}_{ab}$.

The  Yukawa lagrangian of leptons can be written according to the lepton mass eigenstates as,
{\small \bea  -\mathcal{L}^{Y}_{\mathrm{lepton}} &=&\frac{m_{l_b}}{v_2}\sqrt{2} \left[\rho^{0} \bar{l}_bP_Rl_b+  U^{*}_{ba}\bar{\nu}_a P_Rl_b\rho^{+} + V^{L*}_{ba}\overline{N}_a P_Rl_b\rho'^{+}+\mathrm{h.c.}   \right]\crn
	&&+\frac{m_{N_a}}{v_3}\sqrt{2} \left[\chi^{0} \bar{N}_aP_RN_a+  V^{L}_{ba}\bar{l}_b P_RN_a\chi^{-}+\mathrm{h.c.}   \right]\crn
	&&+ \frac{m_{\nu_a}}{v_1}\left[S_1\overline{\nu_{a}}P_L\nu_{b}+\fr{1}{\sqrt{2}} \eta^{+}\left(U^{*}_{ba} \overline{\nu_{a}}P_Ll_{b}+ U_{ba} \overline{l^c_{b}}P_L\nu_{a}\right)+\mathrm{h.c.} \right]\crn
& \supset & \frac{1}{\sqrt{2}} U^{*}_{ba}\bar{\nu}_a\left(\frac{m_{l_b}}{v_2}c_{12}P_R+ \frac{m_{\nu_a}}{v_1}s_{12}P_L\right)l_bH_1^{+}+ \frac{1}{\sqrt{2}}V^{L*}_{ba}\bar{N}_a \left(\fr{m_{l_b}}{v_2}c_{23} P_R+\fr{m_{N_a}}{v_3}s_{23} P_L\right)l_bH_2^{+}  \crn
&& \frac{m_{l_b}}{v_2}\left(-s_\alpha h^0_1+c_\alpha h^0_2\right) \bar{l}_bP_Rl_b +\frac{m_{N_a}}{v_3}h^{0}_3 \bar{N}_aP_RN_a+ \frac{m_{\nu_a}}{v_1}\left(c_\alpha h^0_1+s_\alpha h^0_2\right) \overline{\nu_{a}}P_L\nu_{b} +\mathrm{h.c}.\label{llh}
\eea}
where we have used the Majorana property of the active neutrinos: $\nu^c_a=\nu_a$ with $a=1,2,3$. Furthermore, using the equality $\overline{l^c_{b}}P_L\nu_{a}=  \overline{\nu_{a}}P_Ll_{b}$ for this case the term relating with $\eta^{\pm}$ in the fourth line of (\ref{llh}) is reduced to $\sqrt{2}H_1^{+} \overline{\nu_{a}}P_Ll_{b}$.

From Eq.(\ref{llh}), we realize that the flavor-diagonal $h_1^0 \rightarrow l^+_a l^-_a$ decays occur naturally at the tree level similar to SM. The difference comes from the region of the parameter space of this decay. We recall that in SM this type couping is defined as $h^0\bar{l}_al_a \sim \frac{im_{l_a}}{v}$, whereas in the 331NL model it is given as $h_1^0\bar{l}_al_a \sim -\frac{im_{l_a}s_\alpha}{v_2}=\frac{im_{l_a}}{v}.\frac{c_{(\beta_{12}+\delta)}}{c_{\beta_{12}}}$. This difference is determined through a coefficient $\frac{c_{(\beta_{12}+\delta)}}{c_{\beta_{12}}}$. It is very small and will suppress in the limit $\delta \rightarrow 0$.

Next, we consider other types of LFV coupling. The kinetic term  of the leptons contain the lepton-lepton-gauge boson couplings, namely
\bea \mathcal{L}^{K}_{\mathrm{lepton}} &=& i\overline{L'_a}\gamma^{\mu}D_{\mu}L'_a\crn
&\supset& \frac{g}{\sqrt{2}}\left[ U^*_{ba}\overline{\nu_a}\gamma^{\mu}P_L l_bW^+_{\mu} +U_{ab}\overline{l}_b\gamma^{\mu}P_L\nu_aW^-_{\mu}\right. \crn &+&\left. V^{L*}_{ba}\overline{N_a}\gamma^{\mu}P_L l_bV^+_{\mu} +V^L_{ab}\overline{l}_b\gamma^{\mu}P_LN_aV^-_{\mu}  \right] . \label{cdelepton}\eea
The interaction of the Higgs bosons with the gauge bosons comes from the kinetic term of the scalar fields.
\bea \mathcal{L}^{K}_{\mathrm{scalar}} &&= i\sum_{\Phi=\eta,\rho,\chi}\left( D_\mu \Phi\right) ^\dagger \left( D_{\mu}\Phi\right) = i\sum_{\Phi=\eta,\rho,\chi}\left(\Phi ^\dagger D^\mu \right)  \left( D_{\mu}\Phi\right),\crn
&& \supset \frac{ig^2}{2}\left\lbrace \left[( \eta^0)^2+( \rho^0)^2\right]W^{+\mu}W_{\mu}^- + \left[( \rho^0)^2+( \chi^0)^2\right]\right\rbrace V^{+\mu}V_{\mu}^- \crn
&&+\frac{ig}{\sqrt{2}}\left\lbrace\left[ ( \partial_\mu {\eta^0})\eta^-+\eta^0(\partial_\mu {\eta^-})+( \partial_\mu {\rho^0})\rho^-+\rho^0(\partial_\mu {\rho^-})\right] W_{\mu}^+\right.\crn
&&+\left.\left[ ( \partial_\mu {\rho^0})\rho^-+\rho^0(\partial_\mu {\rho^-})+(\partial_\mu {\chi^0})\chi^-+\chi^0(\partial_\mu {\chi^-})\right]\right\rbrace  V_{\mu}^+ + \mathrm{h.c}. \label{coupHiggs-gauge}\eea
In Eq.(\ref{coupHiggs-gauge}), we obtain couplings of SM-like Higgs boson with charged gauge bosons and charged Higgs bosons. In particular, from the interactions of charged Higgs with $W$-boson and $Z$-boson mentioned as Refs.\cite{ATLAS:2015edr, CMS:2015lsf}, we find out that in this model only $H_1^\pm W^\mp Z$ is non-zero and $H_2^\pm W^\mp Z$ is suppressed. This results in $m_{H_1^\pm}$ being limited to around $600~\mathrm{GeV}$ \cite{ATLAS:2015edr} or around $1000~\mathrm{GeV}$ \cite{CMS:2015lsf}.

It should also be recalled that from the first term in the second line of Eq.(\ref{coupHiggs-gauge}), we can obtain a familiar coupling, which is also present in SM, as $W^{\mu+}W_\mu^{-}h_1^0 \sim igm_W\left(  c_\al s_{12}- s_\al c_{12}\right)=igm_Wc_\delta$. The difference from SM is expressed through the coefficient $c_\delta$, it is also easy to see that the difference disappears when $\delta \rightarrow 0$.

From the above expansions, we show the lepton-flavor-violating couplings of this model in Tab.\ref{albga}.
%%%%%%%%%%%%%%
\begin{table}[h]
	\scalebox{0.82}{
		\begin{tabular}{|c|c|c|c|}
			\hline
			Vertex & Coupling & Vertex &Coupling \\
			\hline
		$\bar{\nu}_al_bH_1^{+}$&$-i\sqrt{2}U^{L*}_{ba}\left( \dfrac{m_{l_b}}{v_2}c_{12}P_R+\dfrac{m_{\nu_a}}{v_1}s_{12}P_L\right)$&$\bar{l}_b\nu_aH_1^{-}$&$-i\sqrt{2}U^{L}_{ab}\left( \dfrac{m_{l_b}}{v_2}c_{12}P_L+\dfrac{m_{\nu_a}}{v_1}s_{12}P_R\right)$\\
		\hline	
			$\bar{N}_a l_bH_2^{+}$ & $-i\sqrt{2}V^{L*}_{ba}\left(\fr{m_{l_b}}{v_2}c_{23} P_R+\fr{m_{N_a}}{v_3}s_{23} P_L\right)$ & $\bar{l}_b N_aH_2^{-}$ &$-i\sqrt{2}V^{L}_{ba}\left(\fr{m_{l_b}}{v_2}c_{23} P_L+\fr{m_{N_a}}{v_3}s_{23} P_R\right)$ \\
			\hline
			$\bar{\nu}_al_bW_\mu^{+}$&$\fr{ig}{\sqrt{2}}U^{L*}_{ba}\ga^\mu P_L$&$\bar{l}_b\nu_aW_\mu^{-}$&$\fr{ig}{\sqrt{2}}U^{L}_{ab}\ga^\mu P_L$\\
			\hline
			$\bar{N}_al_bV_\mu^{+}$&$\fr{ig}{\sqrt{2}}V^{L*}_{ba}\ga^\mu P_L$&$\bar{l}_bN_aV_\mu^{-}$&$\fr{ig}{\sqrt{2}}V^{L}_{ab}\ga^\mu P_L$\\
			\hline
			$\bar{l}_al_ah_1^0$,\,$\bar{l}_al_ah_2^0$&$-\fr{im_{l_a}}{v_2}s_\al$,\,$\fr{im_{l_a}}{v_2}c_\al$&$\bar{\nu}_a\nu_ah_1^0$,\,$\bar{\nu}_a\nu_ah_2^0$&$\fr{im_{\nu_a}c_\al}{v_1}$,\,$\fr{im_{\nu_a}s_\al}{v_1}$\\
			\hline
			$W^{\mu+}W_\mu^{-}h_1^0$&$igm_W\left(  c_\al s_{12}- s_\al c_{12}\right)  $&$V^{\mu+}V_\mu^{-}h_1^0$&$-igm_V s_\al s_{23}$\\
			%%%%%%%%%%%%%%%%%
			\hline
			$h_1^0H_1^{+}W^{\mu-}$&$\dfrac{ig}{2}\left(c_\al c_{12}+s_\al s_{12} \right) (p_{h_1^0}-p_{H_1^{+}})_\mu$&$h_1^0H_1^{-}W^{\mu+}$&$\dfrac{ig}{2}\left(c_\al c_{12}+s_\al s_{12} \right)(p_{H_1^{-}}-p_{h_1^0})_\mu$ \\
			\hline
			$h_1^0H_2^{+}V^{\mu-}$&$\dfrac{ig}{2}s_\al c_{23}(p_{h_1^0}-p_{H_2^{+}})_\mu$&$h_1^0H_2^{-}V^{\mu+}$&$\dfrac{ig}{2}s_\al c_{23}(p_{H_2^{-}}-p_{h_1^0})_\mu$ \\
			\hline
			$W^{\mu+}W_\mu^{-}h_2^0$&$igm_W\left(  c_\al c_{12}+ s_\al s_{12}\right)  $&$V^{\mu+}V_\mu^{-}h_2^0$&$-igm_V c_\al s_{23}$\\
			\hline
			$h_2^0H_1^{+}W^{\mu-}$&$-\dfrac{ig}{2}\left(c_\al s_{12}-s_\al c_{12} \right) (p_{h_1^0}-p_{H_1^{+}})_\mu$&$h_2^0H_1^{-}W^{\mu+}$&$-\dfrac{ig}{2}\left(c_\al s_{12}-s_\al c_{12} \right)(p_{H_1^{-}}-p_{h_1^0})_\mu$ \\
			\hline
			$h_2^0H_2^{+}V^{\mu-}$&$-\dfrac{ig}{2}c_\al c_{23}(p_{h_2^0}-p_{H_2^{+}})_\mu$&$h_2^0H_2^{-}V^{\mu+}$&$-\dfrac{ig}{2}c_\al c_{23}(p_{H_2^{-}}-p_{h_1^0})_\mu$ \\
			\hline
			$\bar{N}_a N_ah_3^0$&$\fr{im_{N_a}}{v_3}$&$V^{\mu+}V_\mu^{-}h_3^0$&$\dfrac{ig^2v_3}{2}$\\
			\hline
			$h_3^0H_2^{+}V^{\mu-}$&$\dfrac{ig}{2}s_{23}(p_{h_1^0}-p_{H_2^{+}})_\mu$&$h_3^0H_2^{-}V^{\mu+}$&$\dfrac{ig}{2}s_{23}(p_{H_2^{-}}-p_{h_1^0})_\mu$ \\
			\hline
			$h_1^0 H_1^{+}H_1^{-}$&$-i \lambda_{h^0_1H_1H_1} $&$h_1^0H_2^{+}H_2^{-}$&$ -i \lambda_{h^0_1H_2H_2}$\\
			\hline
			$h_2^0 H_1^{+}H_1^{-}$&$-i \lambda_{h^0_2H_1H_1} $&$h_2^0H_2^{+}H_2^{-}$&$ -i \lambda_{h^0_2H_2H_2}$\\
			\hline
			$h_3^0 H_1^{+}H_1^{-}$&$-i \lambda_{h^0_3H_1H_1} $&$h_3^0H_2^{+}H_2^{-}$&$ -i \lambda_{h^0_3H_2H_2}$\\
			\hline
	\end{tabular}}
	\caption{Couplings relating with LFV of CP-even Higgs bosons decays in the 331NL model. All the couplings were only considered in the unitary gauge.} \label{albga}
\end{table}
%%%%%%%%%%%%%%%%

where self-couplings of Higgs bosons are determined by Higgs potential as mentioned in Eq.(\ref{potential}), namely  
\bea
\lambda_{h^0_1H_1H_1}&&=v\left[ \left(s_{12}^3c_\al -c_{12}^3s_\al \right)  \left(\lambda_{12} +\tilde{\lambda}_{12}\right) -s_{12}^2 c_{12} s_\al\left( 2\lambda_{2}+\tilde{\lambda}_{12}\right) +  c_{12}^2 s_{12} c_\al\left( 2\lambda_{1}+\tilde{\lambda}_{12}\right) \right],\crn
\lambda_{h^0_1H_2H_2}&&=v\left[ c_{23}^2\left(s_{12} c_{\alpha } \lambda _{12}-2 c_{12} s_{\alpha } \lambda _2\right) +s_{23}^2\left(s_{12} c_{\alpha } \lambda _{13}-c_{12} s_{\alpha } \left(\lambda _{23}+\tilde{\lambda }_{23}\right)\right)\right] ,      \crn
\lambda_{h^0_2H_1H_1}&&=v\left[  \left(s_{\alpha}s_{12}^3+c_{\alpha}c_{12}^3\right)  \left(\lambda _{12}+\tilde{\lambda }_{12}\right) s_{12}^3+c_{\alpha } c_{12} \left(2 \lambda _2+\tilde{\lambda }_{12}\right) s_{12}^2+c_{12}^2
s_{\alpha } \left(2 \lambda _1+\tilde{\lambda }_{12}\right) s_{12}\right] ,\crn
\lambda_{h^0_2H_2H_2}&&= v\left[  \left(2 c_{\alpha } c_{12} \lambda _2+s_{12} s_{\alpha } \lambda _{12}\right) c_{23}^2 +s_{23}^2 \left(s_{12} s_{\alpha } \lambda _{13}+c_{\alpha } c_{12} \left(\lambda _{23}+\tilde{\lambda }_{23}\right)\right)\right] \crn
&&+s_{23} v_3 \left(2 f s_{\alpha }+c_{\alpha } \tilde{\lambda }_{23}\right)
c_{23}   \crn
\lambda_{h^0_3H_1H_1}&&=  v_3 \left(\lambda _{23} s_{12}^2+2 f s_{12} c_{12}+c_{12}^2 \lambda _{13}\right)  \crn
\lambda_{h^0_3H_2H_2}&&= v_3 \left[ \left(\lambda _{23}+\tilde{\lambda }_{23}\right) c_{23}^2+2 s_{23}^2 \lambda _3\right]  +c_{12} s_{23} v \tilde{\lambda }_{23} c_{23}.  
\eea

These couplings are very important ingredients for us to give the analysis for  lepton-flavor-violating decays following.

%%%%%%%%%%%

\section{Analytic formulas for contributions to $\mathrm{H} \rightarrow l_i l_j$}
\label{Analytic}
By conditioning the relationship between the self-coupling constants in the Higgs potential ($f=\lambda_{13}t_{12}=\frac{\lambda_{23}}{t_{12}}$) \cite{Okada:2016whh} and using the rotation relating to $\delta$-parametter of all 2HDM models \cite{Hung:2019jue,Okada:2016whh}, we have identified $h^0_1$ with the SM-like Higgs boson. However, these also leads to the consequences that some couplings such as: $h^0_1\overline{N}_aN_a$, $h^0_1H_1^\pm H_2^\mp$, $h^0_1H_1^\pm V^\mp$, $h^0_1H_2^\pm W^\mp$ are canceled out. Another feature that should also be noted is that W-boson only couple with active neutrinos while V-boson combines all neutral leptons. The couplings of  two remaining CP-even Higgs bosons, $h^0_2$ and $h^0_3$, are also produced based on the conditions imposed on $h^0_1$.

 Firstly, we recall cLFV decay. This is decay channel that has the most  stringent experimental limit and it is often used to constrain other LFV processes. Based on Ref.~\cite{Crivellin:2018qmi}, the total branching ratios   of the cLFV  processes  are
\begin{equation}\label{eq_Gaebaga}
\mathrm{Br}^{Total}(l_i\rightarrow l_j\gamma)\simeq \frac{48\pi^2}{ G_F^2} \left( \left|\mathcal{D}_{(ij)R}\right|^2 +\left|\mathcal{D}_{(ji)L}\right|^2\right) \mathrm{Br}(l_i\rightarrow l_j\overline{\nu_j}\nu_i),
\end{equation}
where $G_F=g^2/(4\sqrt{2}m_W^2)$, and for different charge lepton decays, we use experimental data $\mathrm{Br}(\mu\rightarrow e\overline{\nu_e}\nu_\mu)=100\%, \mathrm{Br}(\tau\rightarrow e\overline{\nu_e}\nu_\tau)=17.82\%, \mathrm{Br}(\tau\rightarrow \mu\overline{\nu_\mu}\nu_\tau)=17.39\% $ as given in Ref.\cite{Patrignani:2016xqp,Tanabashi:2018oca,Zyla:2020zbs}. $\mathcal{D}_{(ij)L,R}$ are factors which were determined based on Feynman diagrams of the cLFV process as shown in Refs.~\cite{Hue:2017lak, Nguyen:2018rlb,Hung:2021fzb,Hue:2021zyw,Hue:2021xap,Hong:2020qxc} for 3-3-1 models. It should be noted that the numerical survey results of this kind were performed as shown in Refs.\cite{Hung:2022kmv, Hung:2022tdj}. Here we will use those results to examine $\mathrm{H} \rightarrow l_i{\bar{l}_j}$ decays.

\begin{figure}[ht]
	\centering
	\begin{tabular}{cc}
		\includegraphics[width=14.0cm]{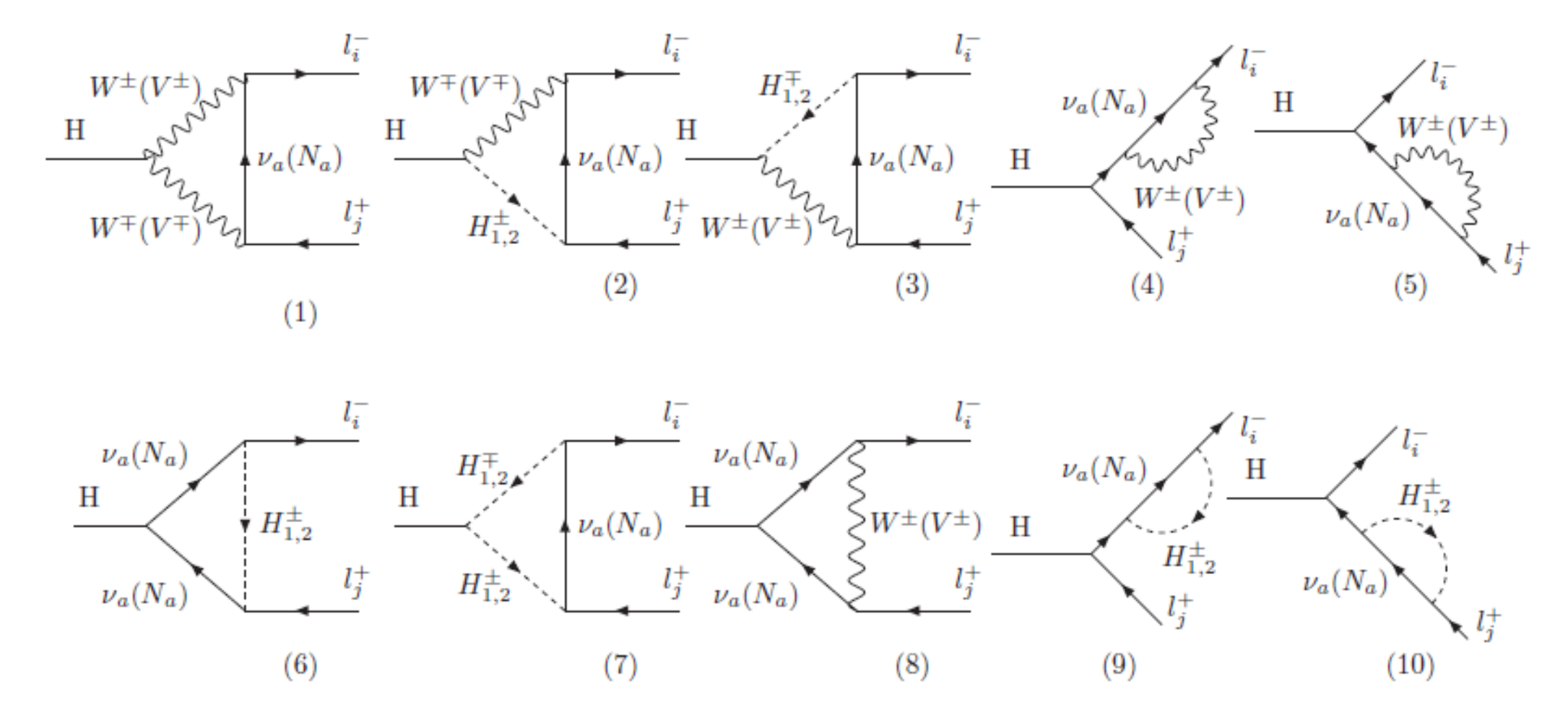} 
	\end{tabular}%
	\caption{ Feynman diagrams at one-loop order of $\mathrm{H} \rightarrow l_i^\pm l_j^\mp$ decays in the unitary gauge, with notation $\mathrm{H} \equiv h^0_1,\,h^0_2,\,h^0_3 $.}
	\label{fig_hmt331}
\end{figure}

For convenience when investigating $\mathrm{H}\rightarrow l_i^{\pm}l_j^{\mp}$ decay, we use scalar factors $\Delta^{(ij)}_{L,R}$. Therefore, the effective Lagrangian of  these decays is
\bea \label{Lag_eff}
 \mathcal{L}_{\mathrm{LFVH}}^\mathrm{eff}= \mathrm{H} \left(\Delta^{(ij)}_L \overline{l_i}P_L l_j +\Delta^{(ij)}_R \overline{l_i}P_R l_j\right) + \mathrm{h.c.}
 \eea
According to the couplings listed in Tab.\ref{albga}, we obtain the one-loop Feynman diagrams contributing to these amplitude in the unitary gauge are shown in Fig.\ref{fig_hmt331}. Inevitably, the scalar factors $\Delta^{(ij)}_{L,R}$  arise from the loop contributions, we only pay attention to all corrections at one-loop order.\\

%%%5%%%%
%%%%%%%%%
%%
The partial width of $\mathrm{H}\rightarrow l_i^{\pm}l_j^{\mp}$ is
\be
\Gamma (\mathrm{H}\rightarrow l_il_j)\equiv\Gamma (\mathrm{H}\rightarrow l_i^{+} l_j^{-})+\Gamma (\mathrm{H} \rightarrow l_i^{-} l_j^{+})
=  \fr{ m_{\mathrm{H}} }{8\pi }\left(\vert \Delta^{(ij)}_L\vert^2+\vert \Delta^{(ij)}_R\vert^2\right), \label{LFVwidth}
\ee
We use the conditions for external momentum as: $p^2_{i,j}=m^2_{i,j}$,\, $(p_i+p_j)^2=m^2_{\mathrm{H}}$ and $m^2_{\mathrm{H}}\gg m^2_{i,j}$,\, this leads to branching ratio of $\mathrm{H}\rightarrow l_i^{\pm}l_j^{\mp}$ decays can be given
\bea  Br(\mathrm{H} \rightarrow l_il_j)=\Gamma (\mathrm{H}\rightarrow l_il_j)/\Gamma^\mathrm{total}_{\mathrm{H}}, \label{brhmt} 
\eea
where $\Gamma^\mathrm{total}_{\mathrm{H}}\simeq 4.1\times 10^{-3}~\mathrm{GeV}$ corresponding to $m_{h^0_1}=125.09\,[\mathrm{GeV}]$ as shown in Refs. \cite{Patrignani:2016xqp,Zyla:2020zbs,CMS:2022ahq,Denner:2011mq}.\\ 
The scalar factors $\Delta^{(ij)}$ are represented as the sum of the contributions of the diagrams in Fig.\ref{fig_hmt331}. We assume that $ \mathcal{M}^{(ij)}_{(k)}$ is the amplitude of the $k^{th}$ diagram and is written in a simplified form,
\be  \mathcal{M}^{(ij)}_{(k)} =f^{(ij)}_{(k)} \times \sum_{a}\mathbb{O}_{ia}\mathbb{O}^*_{ja} \left[ \left(\overline{u_i} P_L v_j\right)   \mathcal{A}^{(k)}_L +  \left( \overline{u_i} P_R v_j\right) \mathcal{ A}^{(k)}_R \right],  \label{Diep} \ee
where, $\mathcal{A}^{(k)}_{L,R}$ are defined in Appendix \ref{appen_loops2},  $f^{(ij)}_{(k)}$ are factors and determined from the vertices in loop of the $k^{th}$ diagram and $\mathbb{O}$ is the notation for the mixing matrix of leptons, it is $U$ for neutrinos and $V$ for neutral leptons. Thus, $\Delta^{(ij)}$ has the form,
\bea
\Delta^{(ij)}=\Delta^{(ij)}_{L}P_L+\Delta^{(ij)}_{R}P_R=\sum_k f^{(ij)}_{(k)} \times \sum_{a}\mathbb{O}_{ia}\mathbb{O}^*_{ja} \left[ \mathcal{A}^{(k)}_LP_L +  \mathcal{ A}^{(k)}_R P_R \right],
\eea
Next, we give the coefficients corresponding to the decays of the CP-even Higgs bosons. There are two separate parts which we denote as $\Delta^{(ij)\nu}_{L,R}$ and $\Delta^{(ij)N}_{L,R}$ corresponding to the contributions of active neutrinos and neutral leptons. These parts are given a specific analytic form based on the couplings in Tab.\ref{albga} and the formulas in Appendix \ref{appen_loops1}. To be specific, we give the results for $h^0_1 \rightarrow l_il_j$ decay here, the rest of the cases are given in Appendix \ref{appen_loops2}. Contributions to $\Delta^{(ij)\nu}_{L,R}$ include all of the diagrams on Fig.\ref{fig_hmt331}. Its analytic form is
%--For active neutrino
\bea  \Delta^{(ij)\nu-h^0_1}_{L,R} &=&  \sum_{a}U_{ia} U_{ja}^{*} \frac{1}{64\pi^2}\left[2g^3(c_{\alpha}s_{12}-s_\alpha c_{12})\times\mathcal{A}^{(1)}_{L,R}(m_{\nu_a},m_W)\right. \crn
&&+g^2(c_{\alpha}c_{12}+s_\alpha s_{12})\times \mathcal{A}^{(2)}_{L,R} (s_{12},c_{12},v_1,v_2,m_{\nu_a},m_W,m_{H^{\pm}_1}) \crn
&&+g^2(c_{\alpha}c_{12}+s_\alpha s_{12})\times \mathcal{A}^{(3)}_{L,R} (s_{12},c_{12},v_1,v_2,m_{\nu_a},m_W,m_{H^{\pm}_1}) \crn
&&+g^3s_{\alpha}\times \mathcal{A}^{(4+5)}_{L,R}(m_{\nu_a},m_W)\crn
&&-2c_{\alpha}\times \mathcal{A}^{(6)}_{L,R}(s_{12},c_{12},v_1,v_2,m_{\nu_a},m_{H^{\pm}_1})\crn
&&-2 \lambda_{h^0H_1H_1}\times \mathcal{A}^{(7)}_{R} (s_{12},c_{12},v_1,v_2,m_{\nu_a},m_{H^{\pm}_1})\crn
&&-(g^3c_{\alpha}/2)\times \mathcal{A}^{(8)}_{L,R}(m_W,m_{\nu_a})\crn
&&+\left.  s_{\alpha} \times \mathcal{A}^{(9+10)}_{L,R} (s_{12},c_{12},v_1,v_2,m_{\nu_a},m_{H^{\pm}_1})\right].  \label{nudeltaL1}\eea
%%%%%%%
Because $h^0_1$ is not couple to neutral leptons and $N_a$ are not also couple to $W^\pm$-bosons and active neutrinos then $\Delta^{(ij)N}_{L,R}$ includes only the contributions of diagrams $(1),\,(2),\,(3),\,(4),\,(5),\,(7),\,(9),\,(10)$ on Fig.\ref{fig_hmt331}. Its representation is
%%%%%%%
\bea  \Delta^{(ij)N-h^0_1}_{L,R} &=& \sum_{a}V_{ia}^LV_{ja}^{L*}  \frac{1}{64 \pi^2}
%---exotic lepton
\left[2g^3s_\al s_{23} \times \mathcal{A}^{(1)}_{L,R}(m_{N_a},m_V) \right.\crn
&&+\left(g^2s_{\alpha}c_{23}\right)\times \mathcal{A}^{(2)}_{L,R} (c_{23},s_{23},v_2,v_3,m_{N_a},m_V,m_{H^{\pm}_2}) \crn
&&+ \left(g^2s_{\alpha}c_{23}\right)\times \mathcal{A}^{(3)}_{L,R} (c_{23},s_{23},v_2,v_3,m_{N_a},m_V,m_{H^{\pm}_2}) \crn
&&+(g^3s_{\alpha})\times \mathcal{A}^{(4+5)}_{L,R}(m_{N_a},m_V)\crn
&&-\left( 2\lambda_{h^0_1H_2H_2}\right) \times \mathcal{A}^{(7)}_{L,R} (c_{23},s_{23},v_2,v_3,m_{N_a},m_{H^{\pm}_2})\crn
&&+\left. (s_{\alpha})\times \mathcal{A}^{(9+10)}_{L,R} (c_{23},s_{23},v_2,v_3,m_{N_a},m_{H^{\pm}_2})\right]  \label{NdeltaL1}\eea
%%%%%%%%%%%%%%%%
In general, the factor for decay $\mathrm{H} \rightarrow l_il_j$ is determined
\bea \label{TotalAmp}
\Delta^{(ij)}_{L,R}=\Delta^{(ij)\nu}_{L,R}+\Delta^{(ij)N}_{L,R}
\eea
%%%%%%%%%%%%%%%%%%%%%
In Appendix \ref{appen_PV}, we have given PV (Passarino-Veltman) functions as the basis for establishing $\Delta^{(ij)}_{L,R}$. The functions that contain divergence are given based on the same technique as mentioned in Ref.\cite{Hue:2017lak} as follows:
\bea  \mathrm{Div}[A_0(M_n)]&=& M_n^2 \Delta_{\epsilon}, \hs  \mathrm{Div}[B^{(n)}_0]= \mathrm{Div}[B^{(12)}_0]= \Delta_{\epsilon}, \crn
\mathrm{Div}[B^{(1)}_1]&=&\mathrm{Div}[B^{(12)}_1] = \frac{1}{2}\Delta_{\epsilon},  \hs  \mathrm{Div}[B^{(2)}_1] = \mathrm{Div}[B^{(12)}_2]= -\frac{1}{2} \Delta_{\epsilon}.  \label{divs1}\eea
We use denotation $ \Delta_{\epsilon}\equiv \frac{1}{\epsilon}+\ln4\pi-\gamma_E $ with $\gamma_E$ is the  Euler constant. Obviously,  
$\Delta^{(ij)}_{L,R}$ contain divergent terms, however, we can separate the divergences and the finite parts in each factor and show that the sum of the divergences is zero. 

Consistent with the formulas in Appendix  B, we give notations to the case of active neutrinos
\bea
a_1 &\rightarrow&  c_{12},\;  a_2\rightarrow s_{12}, \; u_1 \equiv v_1 = \fr{2m_W}{g}s_{12}, \; u_2 \equiv v_2 = \fr{2m_W}{g}c_{12}, \crn
\fr{a_1}{u_1} &=& \fr{g}{2m_W}\fr{c_{12}}{s_{12}}, \; \fr{a_2}{u_2}=\fr{g}{2m_W}\fr{s_{12}}{c_{12}}, \; \fr{a_1a_2}{u_1u_2}=\fr{g^2}{4m^2_W}.\label{aijnulepton}
\eea
we list the non-zero divergent terms of  the relevant diagrams as follows
\bea
\mathrm{Div}\left[\Delta^{(ij)}_{(1)} \right]&=&\mathcal{B}\times (-3)\left( c_\al s_{12}-s_\al c_{12}\right) ,\crn
\mathrm{Div}\left[\Delta^{(ij)}_{(2)} \right]&=&\mathcal{B}\times \frac{3}{2}\left( c_\al c_{12}+s_\al s_{12}\right)\frac{s_{12}}{c_{12}},\crn
\mathrm{Div}\left[\Delta^{(ij)}_{(3)} \right]&=&\mathcal{B}\times \frac{-3}{2}\left( c_\al c_{12}+s_\al s_{12}\right)\frac{c_{12}}{s_{12}},\crn
\mathrm{Div}\left[\Delta^{(ij)}_{(4)} \right]&=&\fr{1}{m_1^2-m_2^2}\left[m_2^2\mathcal{B}_L+m_1^2\mathcal{B}_R\right]\times (\frac{s_\al}{s_{12}}),\crn
\mathrm{Div}\left[\Delta^{(ij)}_{(5)} \right]&=&\fr{1}{m_1^2-m_2^2}\left[m_2^2\mathcal{B}_L+m_1^2\mathcal{B}_R\right]\times (\frac{-s_\al}{s_{12}}),\crn
\mathrm{Div}\left[\Delta^{(ij)}_{(6+8)}\right]&=&\mathcal{B}\times (\frac{3c_\al}{2s_{12}}),\crn
\mathrm{Div}\left[\Delta^{(ij)}_{(9+10)} \right]&=&\mathcal{B}\times (-\frac{3s_\al}{2c_{12}}), \label{canceldiv1} \eea
where
\bea
\mathcal{B}&=&\fr{g^3}{128\pi^2}\fr{m^2_{\nu_{a}}}{m_W^3}\times \Delta_\epsilon\times \left[ \bar{u}_1P_Lv_2\times m_1+\bar{u}_1P_Rv_2\times m_2\right]\crn
\mathcal{B}_L&=&\fr{g^3}{128\pi^2}\fr{m^2_{\nu_{a}}}{m_W^3}\times \Delta_\epsilon\times\bar{u}_1P_Lv_2\times m_1, \;
\mathcal{B}_R =\fr{g^3}{128\pi^2}\fr{m^2_{\nu_{a}}}{m_W^3}\times \Delta_\epsilon\times\bar{u}_1P_Rv_2\times m_2.\label{B_nu}
\eea
We see that sum of all divergence terms is eliminated in the following way:
\bea
&&\mathrm{Div}\left[\Delta^{(ij)}_{(4)} \right]+\mathrm{Div}\left[\Delta^{(ij)}_{(5)} \right]=0,\crn
&&\mathrm{Div}\left[\Delta^{(ij)}_{(1)} \right]+\mathrm{Div}\left[\Delta^{(ij)}_{(2)} \right]+\mathrm{Div}\left[\Delta^{(ij)}_{(3)} \right]+\mathrm{Div}\left[\Delta^{(ij)}_{(6+8)} \right]+\mathrm{Div}\left[\Delta^{(ij)}_{(9+10)} \right]=0. \label{Div_cancel1}
\eea
%
%%%%%%%%%%%%%%%%%%%%%

To apply formulas in Appendix  B in the case of neutral leptons, we also give notations below
%%%%%%%%%%%%%%
\bea
a_1 &\rightarrow&  c_{23},\;  a_2\rightarrow s_{23}, \; u_1 \equiv v_2 = \fr{2m_V}{g}s_{23}, \;  u_2 \equiv v_3 = \fr{2m_V}{g}c_{23}, \crn
\fr{a_1}{u_1} &=& \fr{g}{2m_V}\fr{c_{23}}{s_{23}}, \; \fr{a_2}{u_2}=\fr{g}{2m_V}\fr{s_{23}}{c_{23}}, \; \fr{a_1a_2}{u_1u_2}=\fr{g^2}{4m^2_V}. \label{aijNlepton}
\eea
The divergences corresponding to the diagrams in Fig.\ref{fig_hmt331} are given as:
\bea \mathrm{Div}\left[\Delta_{(1)}^{(ij)}\right] &=& B\times (-3)s_{\alpha}s_{23}, \crn   \mathrm{Div}\left[\Delta_{(2+3)}^{(ij)}\right]&=&  B\times s_{\alpha}\left(3s_{23} -\frac{2}{s_{23}}\right), \crn
\mathrm{Div}\left[\Delta^{(ij)}_{(4)} \right]&=&\fr{1}{m_1^2-m_2^2}\left[m_2^2B_L+m_1^2B_R\right]\times \fr{-s_\al}{s_{23}},\crn
\mathrm{Div}\left[\Delta^{(ij)}_{(5)} \right]&=&\fr{1}{m_1^2-m_2^2}\left[m_2^2B_L+m_1^2B_R\right]\times \fr{s_\al}{s_{23}},\crn
\mathrm{Div}\left[\Delta_{(9+10)}^{(ij)}\right]&=& B\times  s_{\alpha}\times\frac{2}{s_{23}}, \label{canceldiv2}\eea
%----
where
\bea
B&=&\fr{g^3}{128\pi^2}\fr{m^2_{N_{a}}}{m_V^3}\times \Delta_\epsilon\times \left[ \bar{u}_1P_Lv_2\times m_1+\bar{u}_1P_Rv_2\times m_2\right]\crn
B_L&=&\fr{g^3}{128\pi^2}\fr{m^2_{N_{a}}}{m_V^3}\times \Delta_\epsilon\times\bar{u}_1P_Lv_2\times m_1, \;
B_R= \fr{g^3}{128\pi^2}\fr{m^2_{N_{a}}}{m_V^3}\times \Delta_\epsilon\times\bar{u}_1P_Rv_2\times m_2.\label{B_Na}
\eea
and we have
\bea
&&\mathrm{Div}\left[\Delta^{(ij)}_{(4)} \right]+\mathrm{Div}\left[\Delta^{(ij)}_{(5)} \right]=0,\crn
&&\mathrm{Div}\left[\Delta^{(ij)}_{(1)} \right]+\mathrm{Div}\left[\Delta^{(ij)}_{(2+3)} \right]+\mathrm{Div}\left[\Delta^{(ij)}_{(9+10)} \right]=0. \label{Div_cancel2}
\eea
It is easy to see that the sum over all factors is zero. Furthermore,  it is interesting to see that the sums of the two parts in diagram $(4)$ and diagram $(5)$ are self-destructing.

%%%%%%%%%%%%
\section{Numerical results}
\label{numerical_results}
\subsection{\label{Numerical1}Setup parameters}  
We use the well-known experimental parameters \cite{Zyla:2020zbs,Patrignani:2016xqp}: 
the charged lepton masses $m_e=5\times 10^{-4}\,\mathrm{GeV}$,\,  $m_\mu=0.105\,\mathrm{GeV}$,\, $m_\tau=1.776\,\mathrm{GeV}$,\, the SM-like Higgs mass $m_{h^0_1}=125.1\,\mathrm{GeV}$,\,  the mass of the W boson $m_W=80.385\,\mathrm{GeV}$  and the gauge coupling of the $SU(2)_L$ symmetry $g \simeq 0.651$.\\
We can give a limit to the mass of the new charged gauge boson ($m_V$) based on condition $m_Z'\geq 4.0\,\mathrm{TeV}$ as given by Refs.\cite{CMS:2018ipm,ATLAS:2019erb}. In this model, we have the relationship of the neutral gauge boson outside the standard model as $m_Z'^2=\frac{g^2v_3^2c_W^2}{3-4s_W^2}$ resulting in $v_3\geq 10.1\, \mathrm{TeV}$. At LHC$@13\mathrm{TeV}$, we can choose $m_V=4.5~\mathrm{TeV}$ as satisfying the above conditions. This value of $m_V$ is very suitable and will be shown in the numerical investigation below. Mixing angle between light VEVs is chosen $\frac{1}{60}\leq t_{12}\leq 3.5$ in accordance with Refs.\cite{Cepeda:2019klc,Hung:2019jue}. However, $\mathrm{H} \rightarrow l_i l_j$ decays in this model depend very little on the change of $t_{12}$, so we choose $t_{12}=0.5$ in the following investigations. Regarding to the $s_\delta$ parameter, this is an important parameter of 2HDM. In section \ref{Couplings}, we have shown that the couplings of $h^0_1$ will be similar to the ones in Standard Model when $s_\delta \rightarrow 0$, combined with the condition to satisfy all 2HDMs then $c_\delta > 0.99$ according to the results shown in Ref.\cite{Kanemura:2018yai}, we choose $\left| s_\delta\right| < 0.14$. With this arange of $\left| s_\delta\right|$, the model under consideration also predicts the existence of large signal of $h_1^0 \rightarrow Z\ga$. This has been detailed as in recent work \cite{Hung:2019jue}. 

The absolute values of
all Yukawa and Higgs self couplings should be choose less than
$\sqrt{4\pi}$ and $4\pi$, respectively. In addition to the parameters that can impose conditions to determine the value domains, we choose the set of free parameters of this model as: $\lambda_1,\,\lambda_2,\,\lambda_3,\, s_\delta,\,m_{N_1},\,m_{N_2}$ and $m_{H^\pm_2}$
\bea \label{de_par}
\tilde{\lambda}_{12}&&=\frac{2m^2_{H_1^\pm}}{v_1^2+v_2^2}-\frac{2\lambda_{13}v_3^2}{v_2^2},\crn
\tilde{\lambda}_{23}&&=\frac{2m^2_{H_2^\pm}}{v_2^2+v_3^2}-\frac{2\lambda_{13}v_1^2}{v_2^2}.
\eea

and $\lambda_{13}$, $\lambda_{23}$, $\lambda_{12}$, $m_{H_1^\pm}$, $m_{h_2^0}$ was given in Eqs.~\eqref{massedcH}, \eqref{eq_la13}, \eqref{eq_la23}, \eqref{eq_la122}.

Regarding to the parameters of active neutrinos we use the recent results of experiment as shown in Refs.~\cite{Patrignani:2016xqp,Tanabashi:2018oca,Zyla:2020zbs}: 
%%%%
$\Delta m_{21}^2=7.55 \times 10^{-5} \mathrm{eV}^2$,\, $\Delta m_{31}^2=2.424 \times 10^{-3} \mathrm{eV}^2$,\, $s_{12}=0.32$,\, $s_{23}=0.547$,\, $s_{13}=0.0216$
%%%%%%%

The mixing matrix of active neutrinos is derived from the $U \equiv U^{PMNS}$ when we ignore a very small deviation \cite{Ibarra:2010xw,upmns,Kuipers:2012rf}. That way, one gives $U=U^L (\theta_{12}^\nu,\theta_{13}^\nu,\theta_{23}^\nu)$ and $U^{\dagger}=U^{L\dagger} (\theta_{12}^\nu,\theta_{13}^\nu,\theta_{23}^\nu)$, with $\theta_{ij}^\nu$ are mixing angles of active neutrinos, the parameterized form of $U$ matrix is
%%%%%%%%55
\bea U^L (\theta_{12},\theta_{13},\theta_{23})&=&\left(
\begin{array}{ccc}
	1 & 0 & 0 \\
	0 & \cos\theta_{23} & \sin\theta_{23} \\
	0 & -\sin\theta_{23} & \cos\theta_{23} \\
\end{array}
\right)
\left(
\begin{array}{ccc}
	\cos\theta_{13} & 0 & \sin\theta_{13} \\
	0 & 1 & 0 \\
	-\sin\theta_{13} & 0 & \cos\theta_{13} \\
\end{array}
\right)\crn
&\times&
\left(
\begin{array}{ccc}
	\cos\theta_{12} & \sin\theta_{12} & 0 \\
	-\sin\theta_{12} & \cos\theta_{12} & 0 \\
	0 & 0 & 1 \\
\end{array}
\right). \label{mixingpar}\eea

%%%%%%%%%%%%

Neutral leptons are also mixed in a common way based on Eq.~(\ref{mixingpar}), by choosing $V^L \equiv U^L(\theta_{12}^N,\theta_{13}^N,\theta_{23}^N)$,  with $\theta_{ij}^N$ are mixing angles of neutral leptons. The  parameterization of $V^L$ is chosen so that the LFV decays can be obtained large signals. According to that criterion, we can give some cases corresponding to large mixing angle of exotic leptons and there are following interesting cases: $V^L \equiv U^L(\frac{\pi}{4},\frac{\pi}{4},\frac{\pi}{4})$,\hs$V^L \equiv U^L(\frac{\pi}{4},\frac{\pi}{4},-\frac{\pi}{4})$ and $V^L \equiv U^L(\frac{\pi}{4},0,0)$. The other cases only change the total amplitude sign of the diagrams in Fig.(\ref{fig_hmt331}) without changing the branching ratio of the LFVHD process.

\subsection{\label{Numerical2} The contributing components to $\mathrm{H} \rightarrow \mu\tau$}  
\begin{figure}[ht]
	\centering
	\begin{tabular}{cc}
		\includegraphics[width=7.0cm]{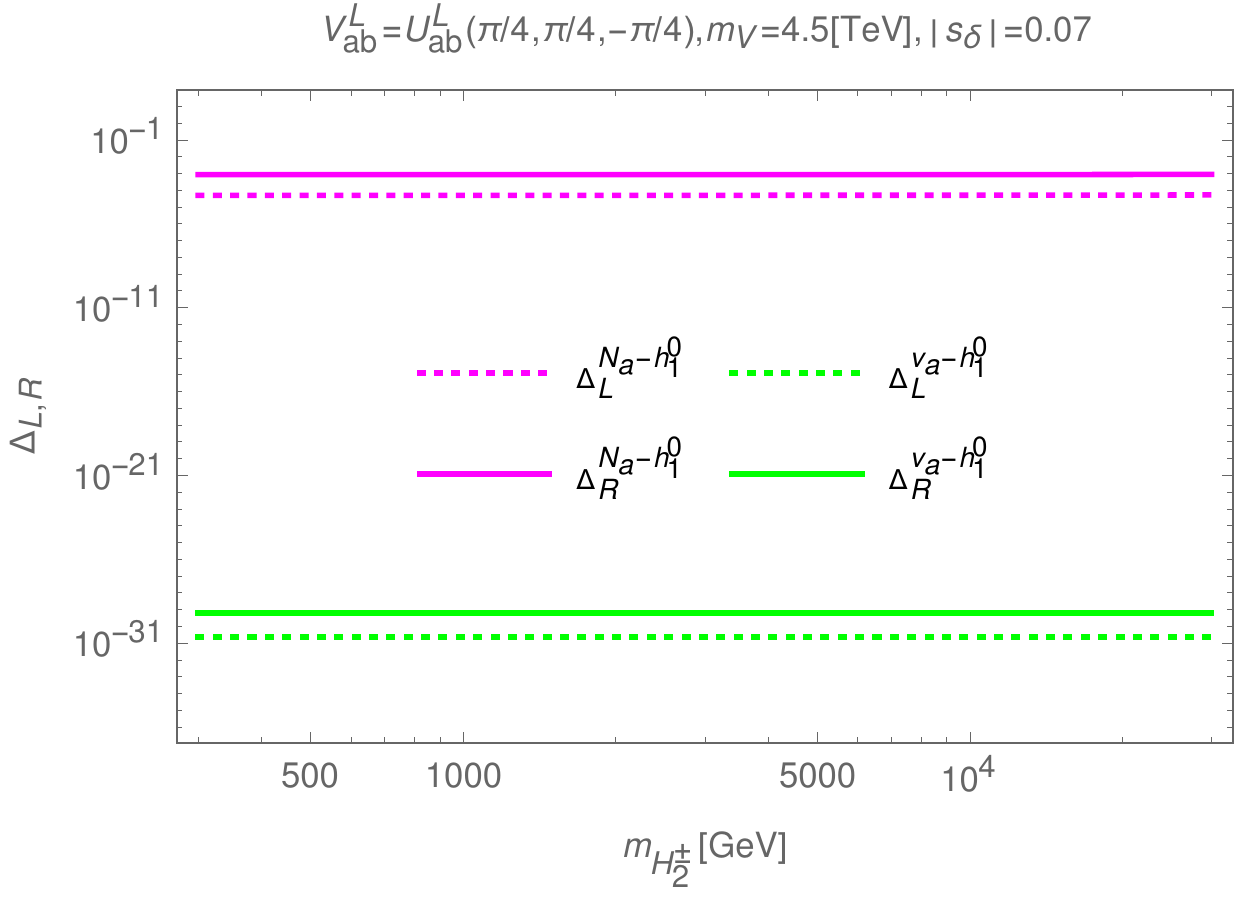}& \includegraphics[width=7.0cm]{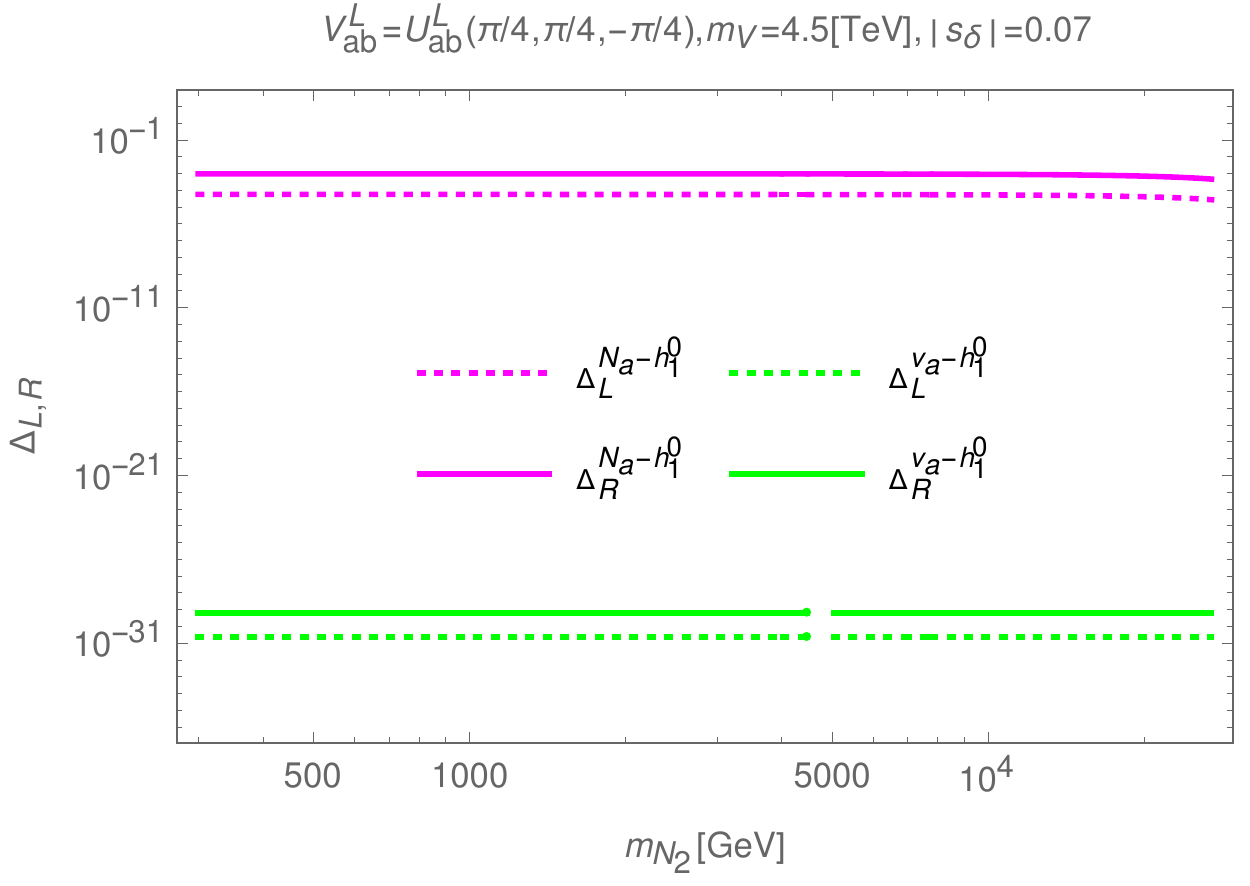}
	\end{tabular}%
	\caption{ Plots of $\Delta^{\nu_a-h^0_1}_{L,R}$ and $\Delta^{N_a-h^0_1}_{L,R}$ depending on $m_{H^\pm_2}$ (left) or  $m_{N_2}$ (right) in the case of $V^L_{ab} = U^L_{ab}(\pi/4,\pi/4,-\pi/4)$.}
	\label{fig_com1}
\end{figure}

Among $\mathrm{H} \rightarrow l_i l_j$, $\mathrm{H} \rightarrow \mu\tau$ usually has the largest decay width because of $m_\tau \gg m_\mu \gg m_e$ as mentioned in some publications such as Refs.\cite{Hung:2022kmv,Hung:2021fzb,Nguyen:2018rlb,Hue:2015fbb,Thuc:2016qva}. Therefore, in the following numerical investigation, we will be interested only in $\mathrm{H} \rightarrow \mu\tau$, other decay channels of this type ($\mathrm{H} \rightarrow e\tau$, $\mathrm{H} \rightarrow e\mu$) can be performed in a similar way.\\
As the results we have shown in Section \ref{Analytic} and Appendix \ref{appen_loops2} , the contribution to $\Delta_{L,R}$ of $\mathrm{H} \rightarrow \mu\tau$ always includes two components $\Delta_{L,R}^{\nu_a}$ and $\Delta_{L,R}^{N_a}$ corresponding to the appearance of active neutrinos and neutral leptons in the diagrams of Fig. \ref{fig_hmt331}. For each of those contributors, we have also shown that terms containing divergences are automatically eliminated as mentioned in Eqs.(\ref{canceldiv1},\ref{canceldiv2}) and Eqs.(\ref{canceldiv3},\ref{canceldiv4},\ref{canceldiv5}). It is mean that  sum of all terms of a component is finite. \\
%%%
 We use the notations, $\Delta^{\nu_a-h^0_i}_{L,R}$ and $\Delta^{N_a-h^0_i}_{L,R}$, for the contributing components of active neutrinos and neutral leptons.These components depend very strongly on the form of the mixing matrix of neutral leptons $V^L_{ab}$. As the results shown in Ref.\cite{Hung:2022kmv}, the signal of $h^0_1 \rightarrow \mu\tau$ in the case of $V^L_{ab} = U^L_{ab}(\pi/4,\pi/4,-\pi/4)$ is the largest. So, we represent the results of $\Delta^{\nu_a-h^0_1}_{L,R}$ and $\Delta^{N_a-h^0_1}_{L,R}$ in Fig.\ref{fig_com1}.
 %%%5%%%%
%%%%%%%%%

\begin{figure}[ht]
	\centering
	\begin{tabular}{cc}
		\includegraphics[width=7.0cm]{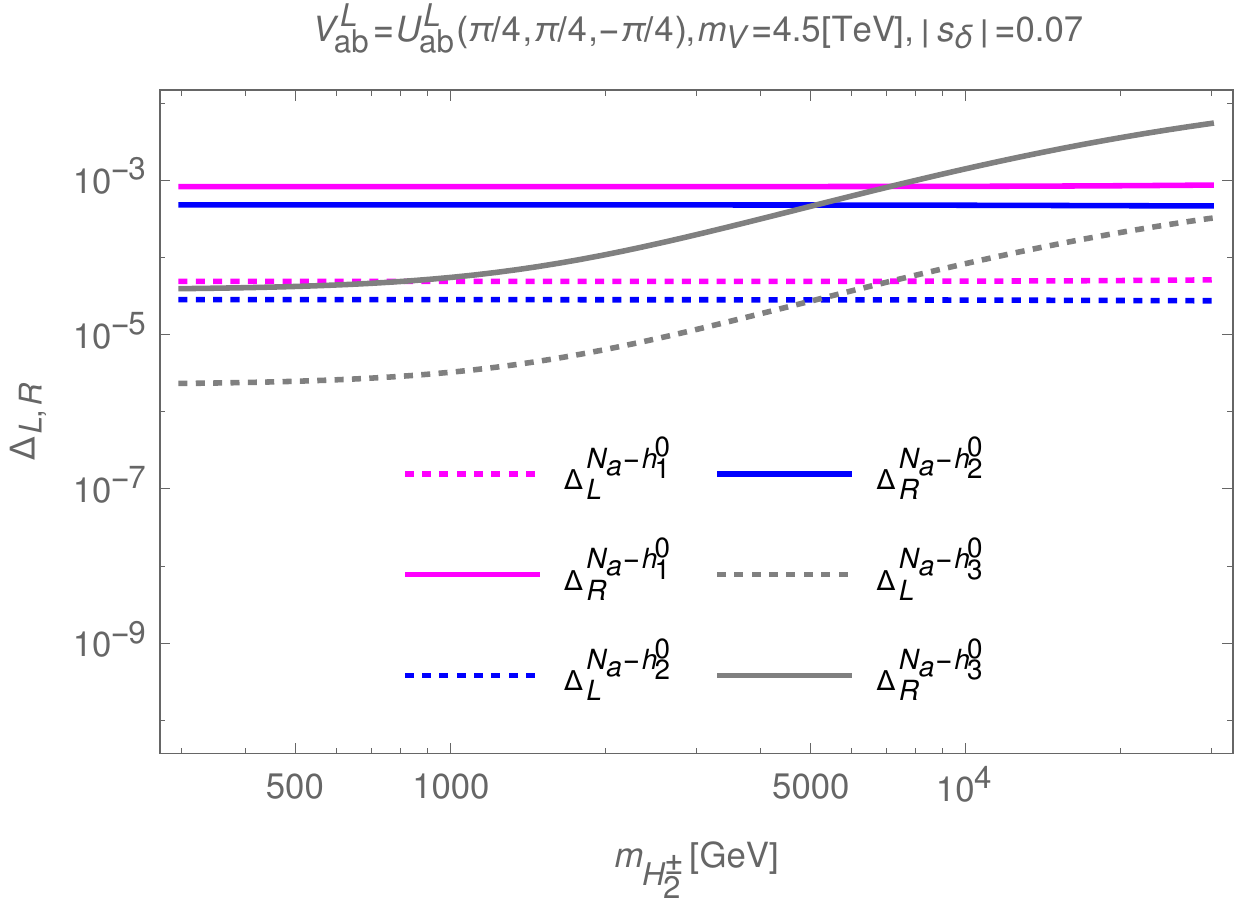}& \includegraphics[width=7.0cm]{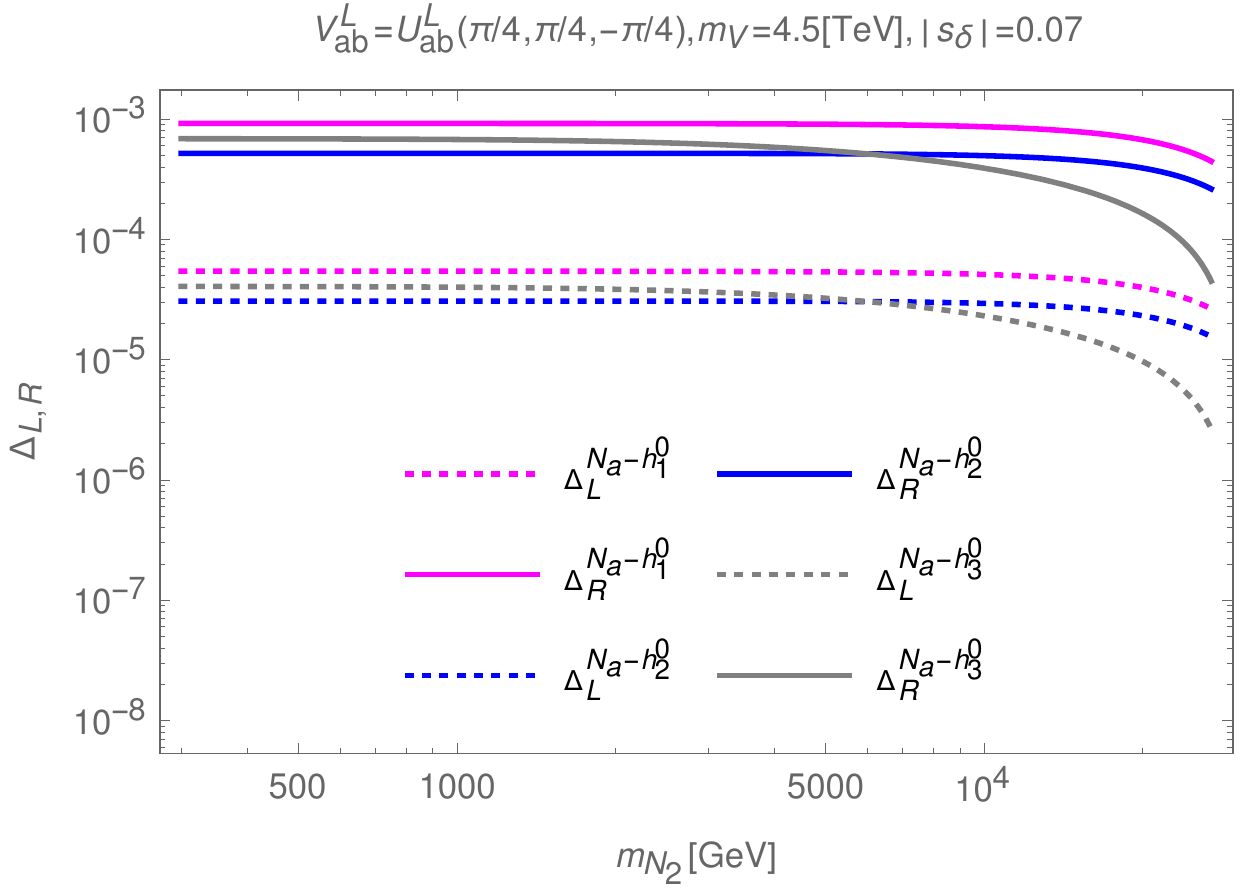}\\
		\includegraphics[width=7.0cm]{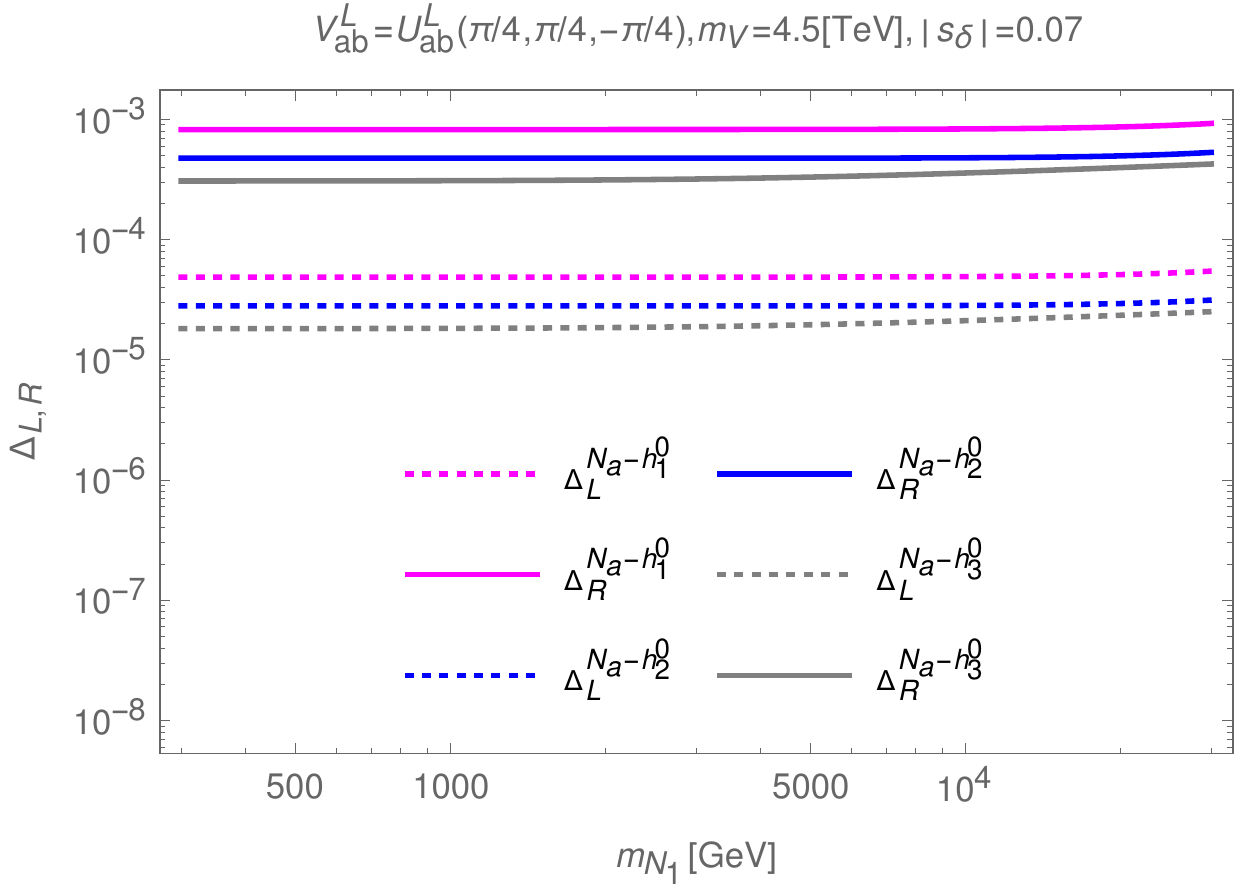}& \includegraphics[width=7.0cm]{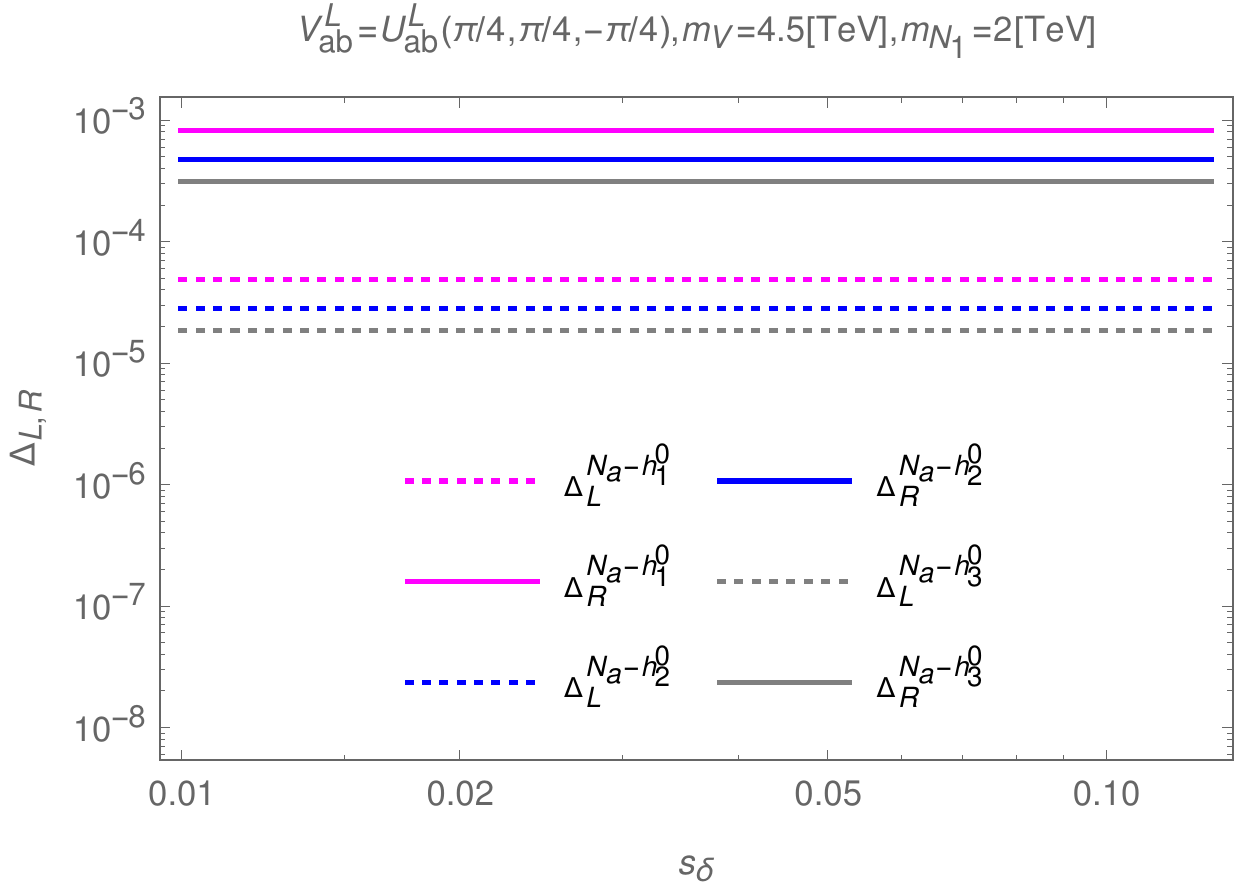}
	\end{tabular}%
	\caption{ Plots of $\Delta^{N_a-h^0_i}_{L,R}$ depending on $m_{H^\pm_2}$, $m_{N_2}$ (first row) or $m_{N_1}$, $s_{\delta}$ (second row) in the case of $V^L_{ab} = U^L_{ab}(\pi/4,\pi/4,-\pi/4)$.}
	\label{fig_c2}
\end{figure}

The numerical investigation in Fig.\ref{fig_com1} show that $\Delta^{\nu_a-h^0_1}_{L,R} \ll \Delta^{N_a-h^0_1}_{L,R} $, leading to the main contribution to $\Delta_{L,R}$ being $\Delta^{N_a-h^0_1}_{L,R}$ and $\Delta^{\nu_a-h^0_1}_{L,R}$ can be omitted in later calculations. This statement is also true for the remaining decays of CP-even Higgs bosons, $\mathrm{H} \rightarrow \mu\tau$. Therefore, in the next survey, we only numerically investigate  with $\Delta^{N_a-h^0_i}_{L,R}$. The results are shown in Fig.\ref{fig_c2}.
%%%5%%%%
%%%%%%%%%

With the change of $m_{H^\pm_2}$, $m_{N_2}$, $m_{N_1}$, $s_{\delta}$ parameters, $\Delta^{N_a-h^0_i}_{R}$ is usually larger than $\Delta^{N_a-h^0_i}_{L}$ with a coefficient proportional to the fraction $\frac{m_\tau}{m_\mu}$. The parameter space regions are chosen so that $\mathrm{Br}(h \rightarrow \mu \tau (Z\gamma))$ can reach the largest value, consistent with the results in Refs.\cite{Hung:2022kmv,Hung:2019jue}. For example, the signal of $\mathrm{Br}(h \rightarrow Z\gamma)$ can achieve $1\leq \mathrm{R}_{Z\ga/\ga\ga}\leq 2$ in Ref.\cite{Hung:2019jue} or the signal of $\mathrm{Br}(h \rightarrow \mu \tau)$ within the limit of $\mathrm{Br}(l_i \rightarrow l_j \gamma)$ can reach about $10^{-4}$ in Ref.\cite{Hung:2022kmv}. In these parameter space regions, $\Delta^{N_a-h^0_i}_{L,R}$ varies rather strongly in the large value domain of $m_{H^\pm_2}$ or  $m_{N_2}$ (first row). Contrary, $\Delta^{N_a-h^0_i}_{L,R}$ depends very little on $m_{N_1}$ or $s_{\delta}$ in the second row of Fig.\ref{fig_c2}. Therefore, in the following content, we only consider the dependence of the partial width of $\mathrm{H}\rightarrow \mu \tau$ on $m_{H^\pm_2}$ or  $m_{N_2}$.\\
From Eq.(\ref{LFVwidth}), we can give a way to determine a coefficients,  $\Delta_L^2+\Delta_R^2=8\pi \times \frac{\mathrm{\Gamma}(\mathrm{H} \rightarrow \mu\tau)}{m_{\mathrm{H}}}\sim \frac{\mathrm{\Gamma}(\mathrm{H} \rightarrow \mu\tau)}{m_{\mathrm{H}}}$, this is a quantity that is proportional to the partial width ($\mathrm{\Gamma}(\mathrm{H} \rightarrow \mu\tau)$) per unit mass of a particle. It has determined based on the diagrams in Fig.\ref{fig_hmt331}. The results in Fig.\ref{fig_c2} show that $\frac{\mathrm{\Gamma}(h^0_1 \rightarrow \mu\tau)}{m_{h^0_1}}$ of SM-like Higgs boson is usually larger than the corresponding ones of two remaining CP-even Higgs bosons. However, there is an important difference, in the parameter space domain where $m_{H^\pm_2}\geq 5\,[\mathrm{TeV}]$ occurs $\frac{\mathrm{\Gamma}(h^0_3 \rightarrow \mu\tau)}{m_{h^0_3}}>\frac{\mathrm{\Gamma}(h^0_1 \rightarrow \mu\tau)}{m_{h^0_1}}$ (the left of the first row in Fig.\ref{fig_c2}). This difference is due to the fact that $h^0_3$ does not couple with particles similar to in SM such as $W^\pm$, $\nu_a$, $l_a$ and the contributions due to coupling with particles outside the SM change very rapidly in the high energy scale.

\subsection{\label{Numerical3} Numerical results of $\mathrm{H} \rightarrow \mu\tau$} 
\begin{figure}[ht]
	\centering
	\begin{tabular}{cc}
		\includegraphics[width=7.0cm]{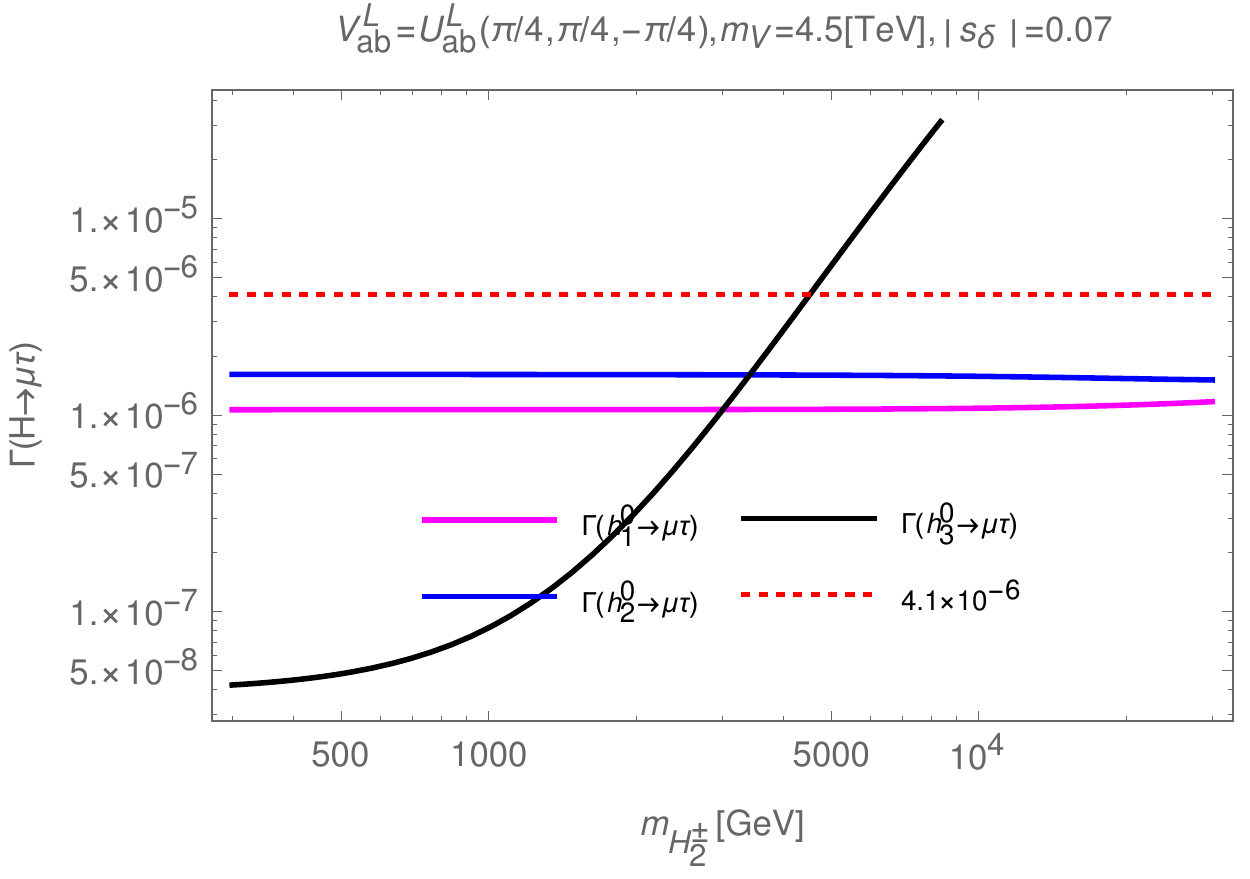}& \includegraphics[width=7.0cm]{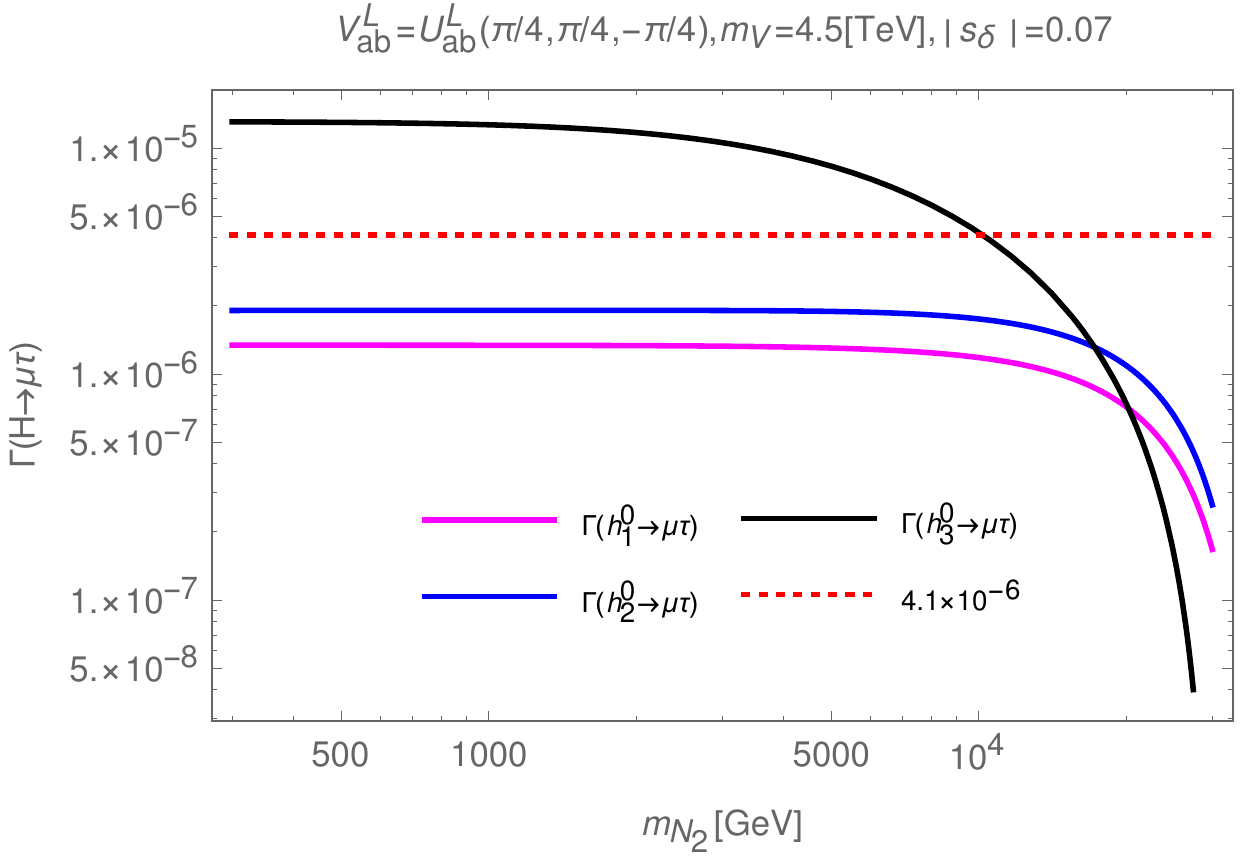}
	\end{tabular}%
	\caption{ Plots of $\mathrm{\Gamma}(h^0_1 \rightarrow \mu\tau)$ depending on $m_{H^\pm_2}$ (left) or  $m_{N_2}$ (right) in the case of $V^L_{ab} = U^L_{ab}(\pi/4,\pi/4,-\pi/4)$.}
	\label{fig_Gam}
\end{figure}

%%%5%%%%
In Sec.\ref{Higgs_spectral}, we have shown the requirements to identify $h^0_1$ with the Higgs boson in the standard model. Thus, among three CP-even Higgs bosons of this model, $h^0_1$ is the lightest, two remaining others, $h^0_2$ and $h^0_3$, have a larger masses. According to Eq.(\ref{LFVwidth}), we obtain the limit $\mathrm{\Gamma}(h^0_1 \rightarrow \mu\tau) \leq 4.1 \times 10^{-6}$ when $m_{h^0_1}=125.09\, [\mathrm{TeV}]$. The numerical results in fig.\ref{fig_c2} also show that $\Delta^{N_a-h^0_i}_{L,R}$ are more dependent on $m_{H^\pm_2}$ or  $m_{N_2}$. Therefore, we will examine the dependence of the partial width of  $\mathrm{H}\rightarrow \mu \tau$ on $m_{H^\pm_2}$ or  $m_{N_2}$ as shown in Fig.  \ref{fig_Gam} where the selected parameter space is guaranteed to be $\mathrm{\Gamma}(h^0_1 \rightarrow \mu\tau) \leq 4.1 \times 10^{-6}$. It means that $\mathrm{Br}(h^0_1 \rightarrow \mu\tau) \leq 10^{-3}$ which is consistent with the current experimental data.
%%%5%%%%
%%%%%%%%%o

The numerical results in Fig.\ref{fig_Gam} show that the dependence of $\mathrm{\Gamma}(h^0_1 \rightarrow \mu\tau)$ on $m_{H^\pm_2}$ or  $m_{N_2}$. In which, the partial decay width of SM-like Higgs boson is $\mathrm{\Gamma}(h^0_1 \rightarrow \mu\tau) \leq 4.1 \times 10^{-6}$, another CP-even Higgs boson, $h^0_2$, also has $\mathrm{\Gamma}(h^0_2 \rightarrow \mu\tau) \leq 4.1 \times 10^{-6}$ and is very close to $\mathrm{\Gamma}(h^0_1 \rightarrow \mu\tau)$. Specially, there is a different behavior from case of $h^0_3$, $\mathrm{\Gamma}(h^0_3 \rightarrow \mu\tau)$ varies a lot in the area of the parameter space under consideration and can receive values larger than $4.1 \times 10^{-6}$. This result is attributed to the fact that $h^0_3$ only interacts with particles outside the SM. \\
The parameter space regions for numerical investigation of LFVHDs are often constrained by the experimental limits of the cLFV decays. Among the cLFV decays, $\mathrm{Br}(\mu \rightarrow e\gamma)$  has the strictest limit \cite{Patrignani:2016xqp,Zyla:2020zbs}. With the model under consideration, one has been shown that in the regions of the parameter space where $\mathrm{Br}(\mu \rightarrow e\gamma)$ satisfies the experimental limit, $\mathrm{Br}(\tau \rightarrow e\gamma)$ and $\mathrm{Br}(\tau \rightarrow \mu\gamma)$ are also satisfied \cite{Hung:2022kmv,Hung:2022tdj}. Therefore, we only need to examine $\mathrm{\Gamma}(\mathrm{H} \rightarrow \mu\tau)$ in the regions where $\mathrm{Br}(\mu \rightarrow e\gamma)$ satisfies the experimental limit.

%%%5%%%%
\begin{figure}[ht]
	\centering
	\begin{tabular}{cc}
		\includegraphics[width=7.0cm]{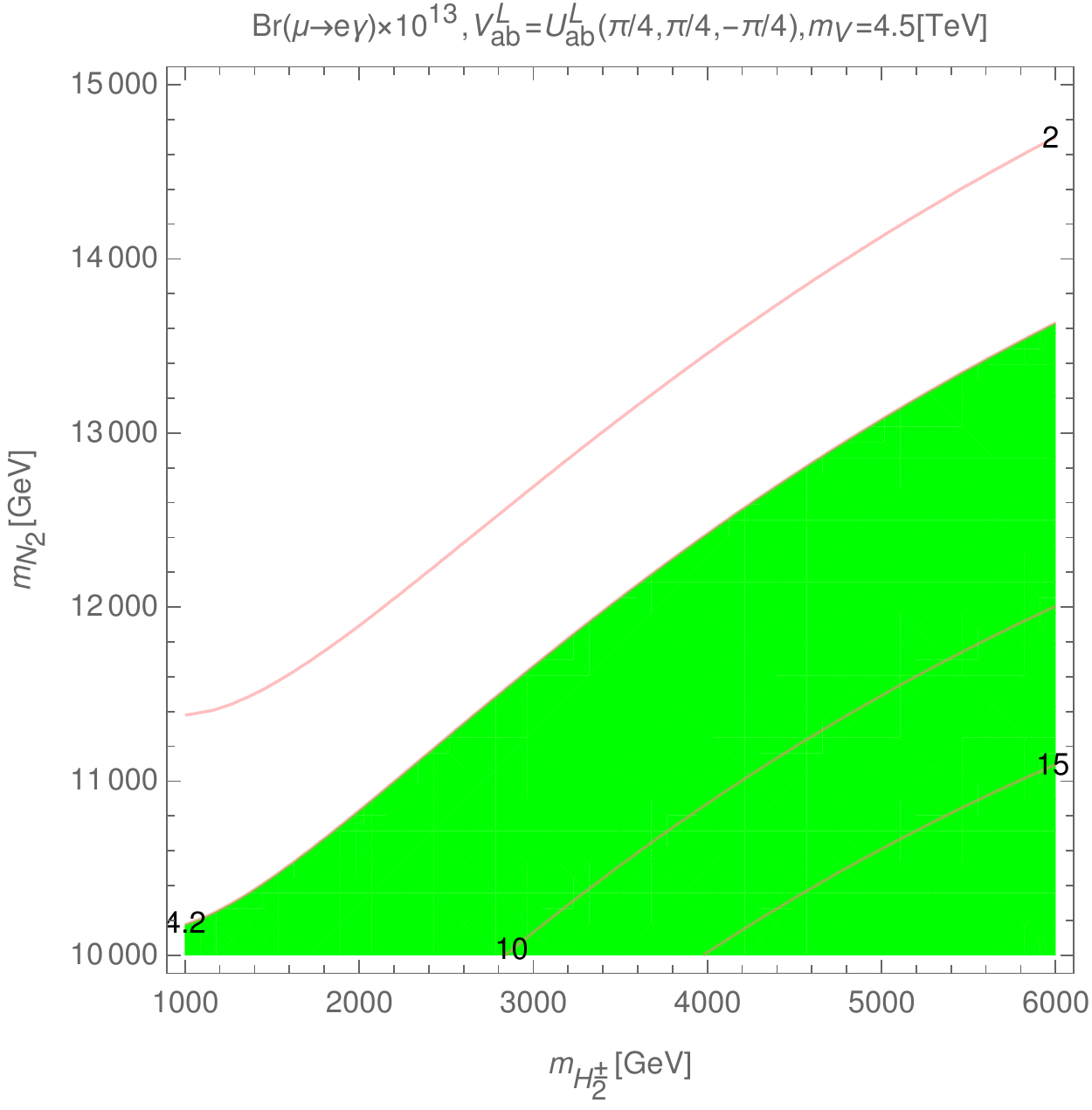}& \includegraphics[width=7.0cm]{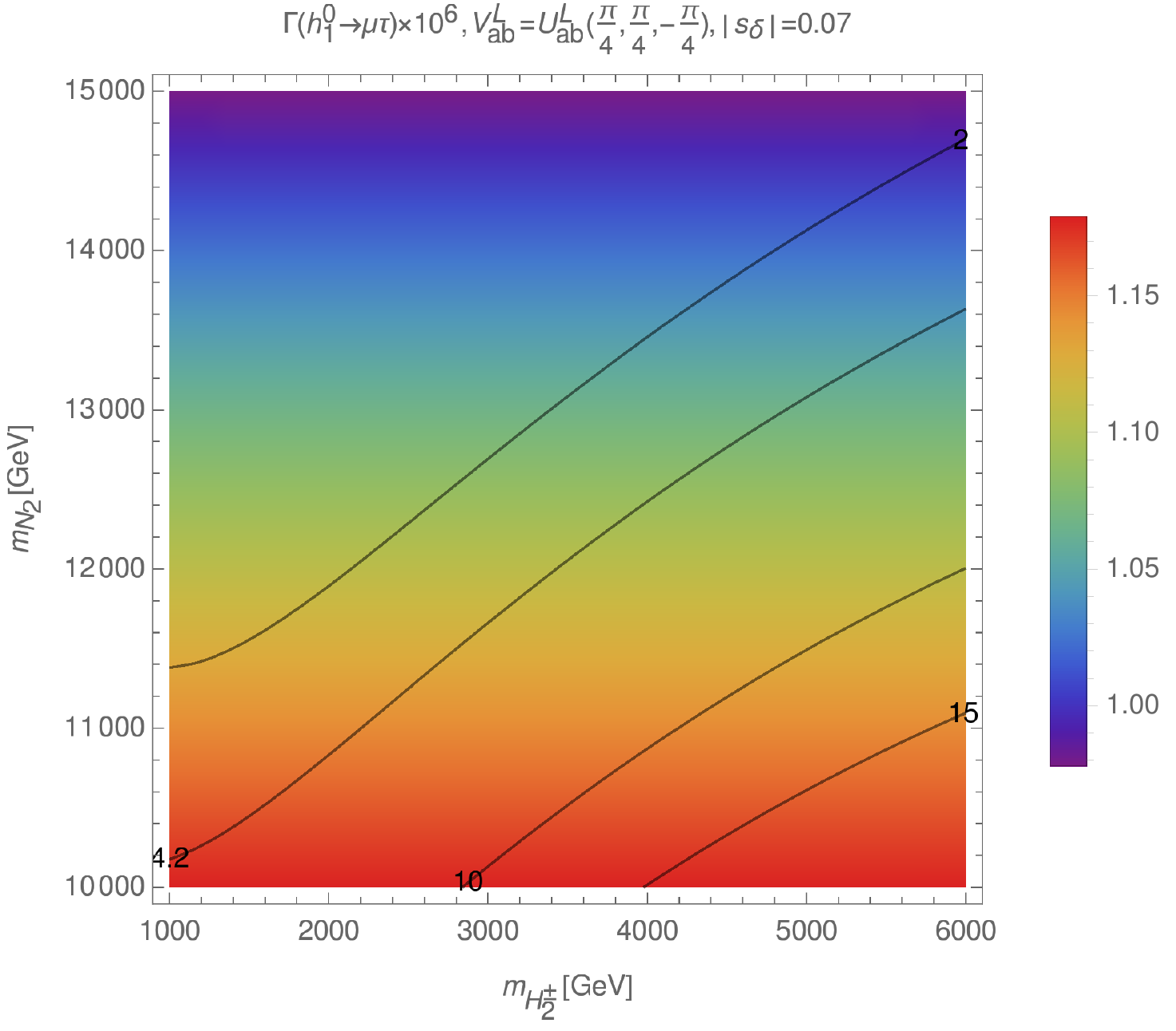}\\
		\includegraphics[width=7.0cm]{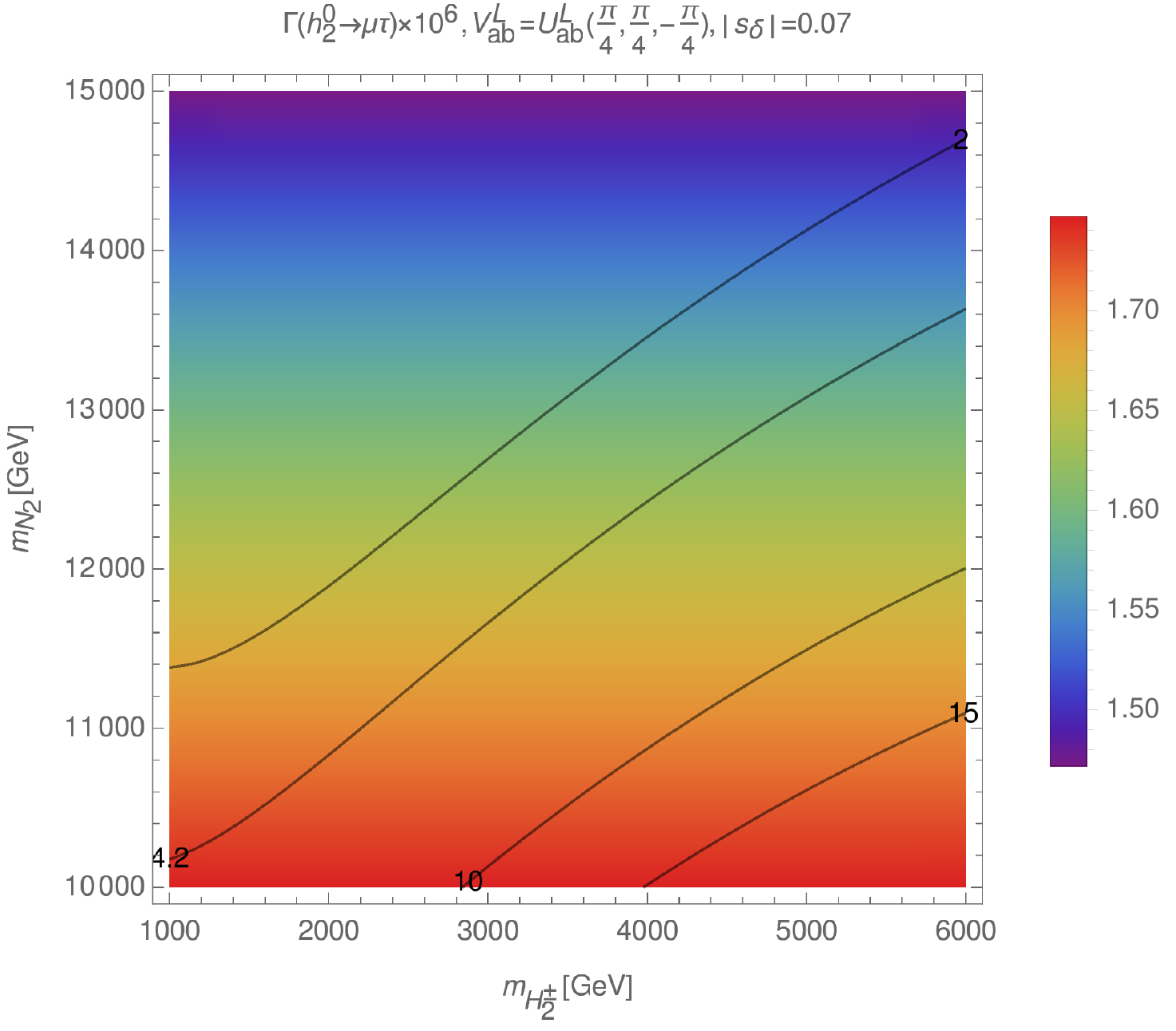}& \includegraphics[width=7.0cm]{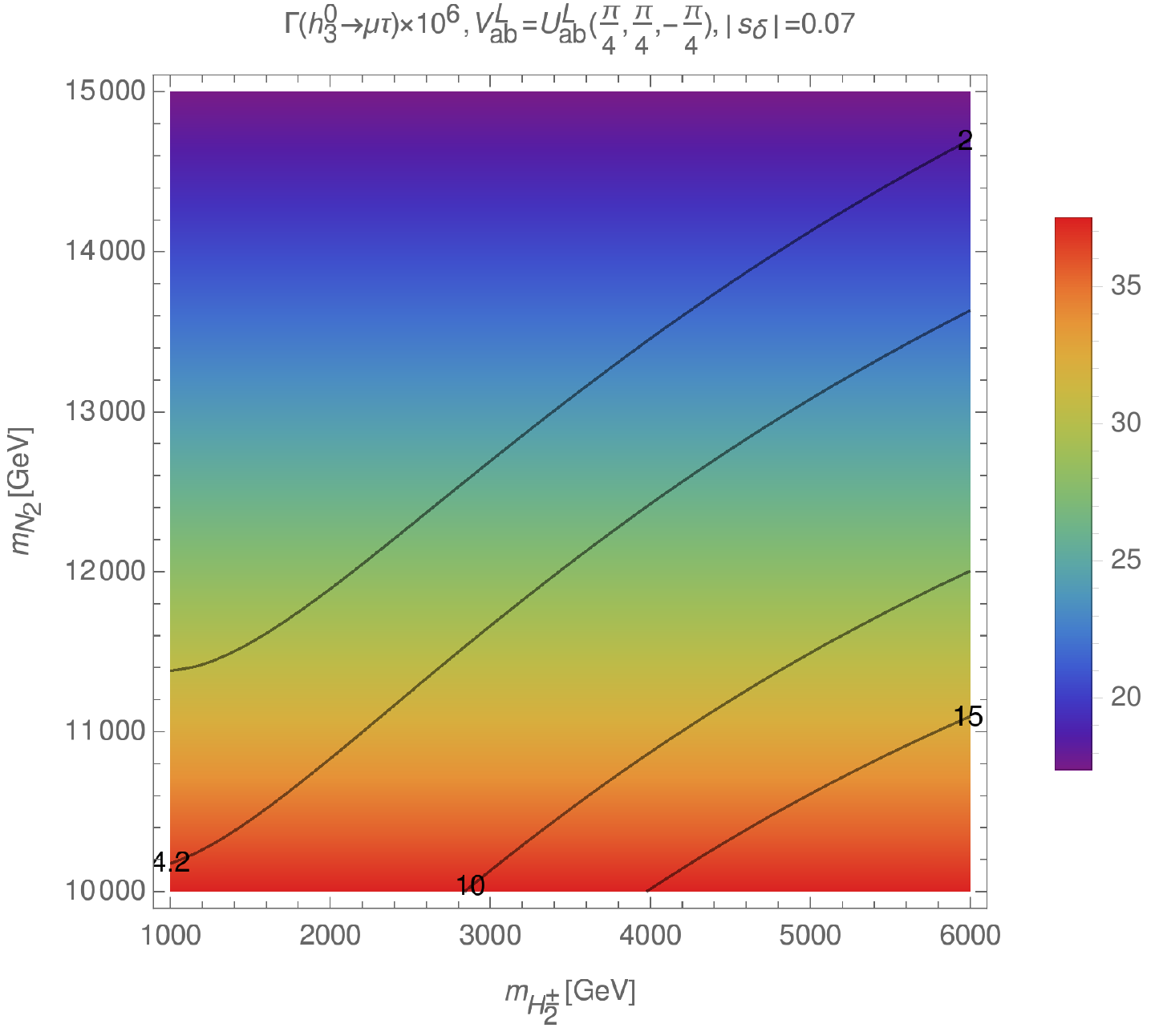}\\
	\end{tabular}%
	\caption{ Contour plots of $\mathrm{Br}(h^0_1 \rightarrow \mu\tau)$ as functions of $m_{H^\pm_2}$ (the left of the first row) and density plots of $\mathrm{\Gamma}(h^0_1 \rightarrow \mu\tau)$ as functions of $m_{H^\pm_2}$ and $m_{N_2}$ in the case of $V^L_{ab} = U^L_{ab}(\pi/4,\pi/4,-\pi/4)$. The black cuvers show the constant values of $\mathrm{Br}(\mu \rightarrow e\gamma)\times 10^{13}$.}
	\label{fig_raint1}
\end{figure}
%%%%%%%%%o

We use the same technique as Ref.\cite{Hung:2022kmv,Hung:2022tdj,Hung:2021fzb} to show the allowed spatial region satisfying the experimental limits of cLFV as the left panel of the first row in Fig.\ref{fig_raint1}. According to this result, the colorless part where $\mathrm{Br}(\mu \rightarrow e\gamma) < 4.2 \times 10^{-13}$ is the allowed region, otherwise the green region is excluded by experimental data. In these allowed spatial regions, the partial width of SM-like Higgs boson is always smaller than the current experimental limit, $\mathrm{\Gamma}(h^0_1 \rightarrow \mu\tau) \leq 4.1 \times 10^{-6}$, and can achieve the largest of about $1.15 \times 10^{-6}$, the corresponding signals of $h^0_2$ and $h^0_3$ are both larger than that of $h^0_1$ as shown in the second row of Fig.\ref{fig_raint1}. This result also helps us predict that the signals of two heavier CP-even Higgs bosons can be found, namely, $\mathrm{\Gamma}(h^0_2 \rightarrow \mu\tau) \leq 1.7 \times 10^{-6}$, $\mathrm{\Gamma}(h^0_3 \rightarrow \mu\tau) \leq 35 \times 10^{-6}$, when $m_{N_2}\sim 10~[TeV]$ and $m_{H^\pm_2}\sim 1~[TeV]$. That implies is the expected things $m_{h^0_2}\sim 185~[GeV]$, $m_{h^0_3}\sim 3800~[GeV]$ can be given out.

%%%5%%%%
\begin{figure}[ht]
	\centering
	\begin{tabular}{cc}
		\includegraphics[width=7.0cm]{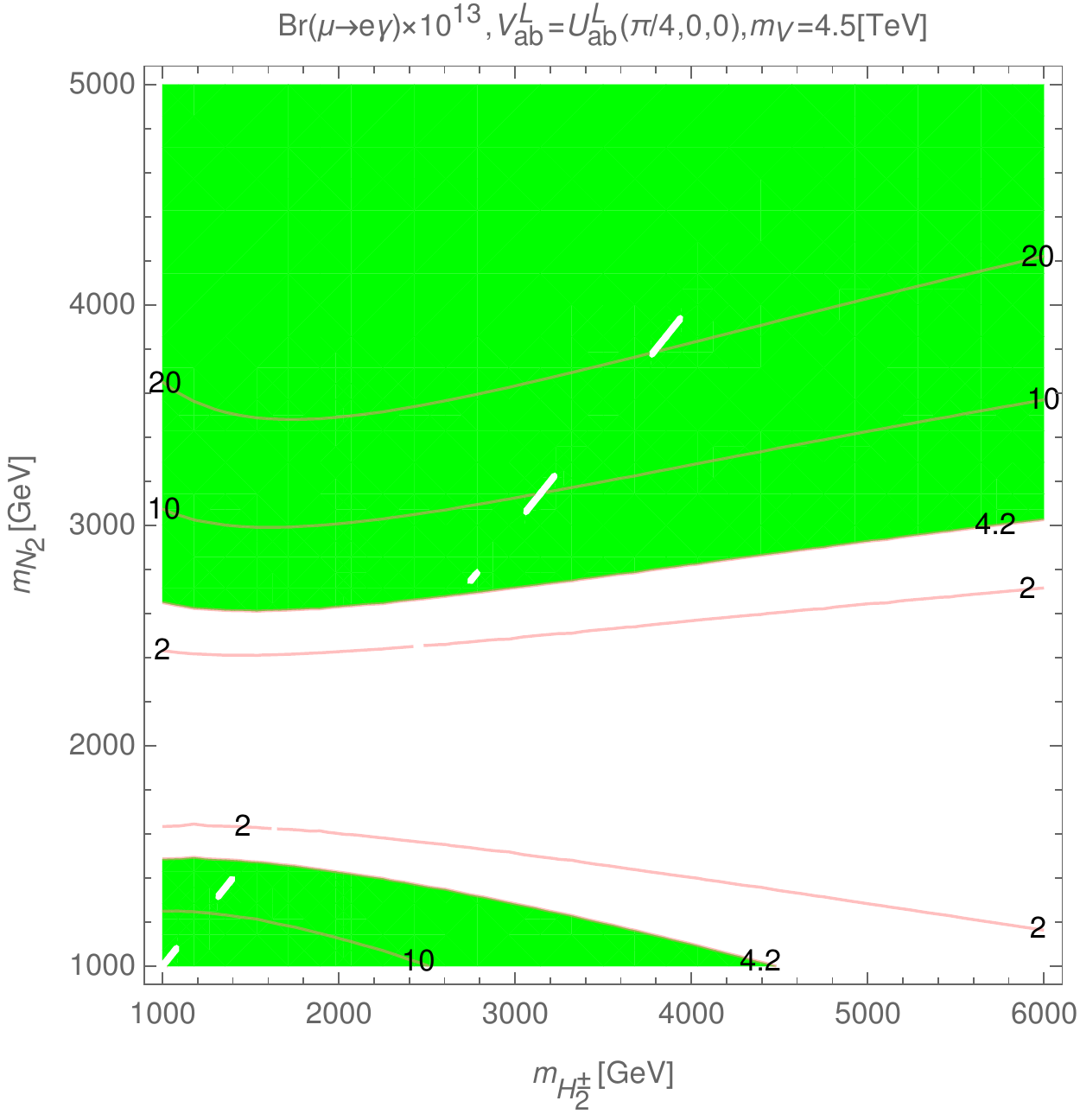}& \includegraphics[width=7.0cm]{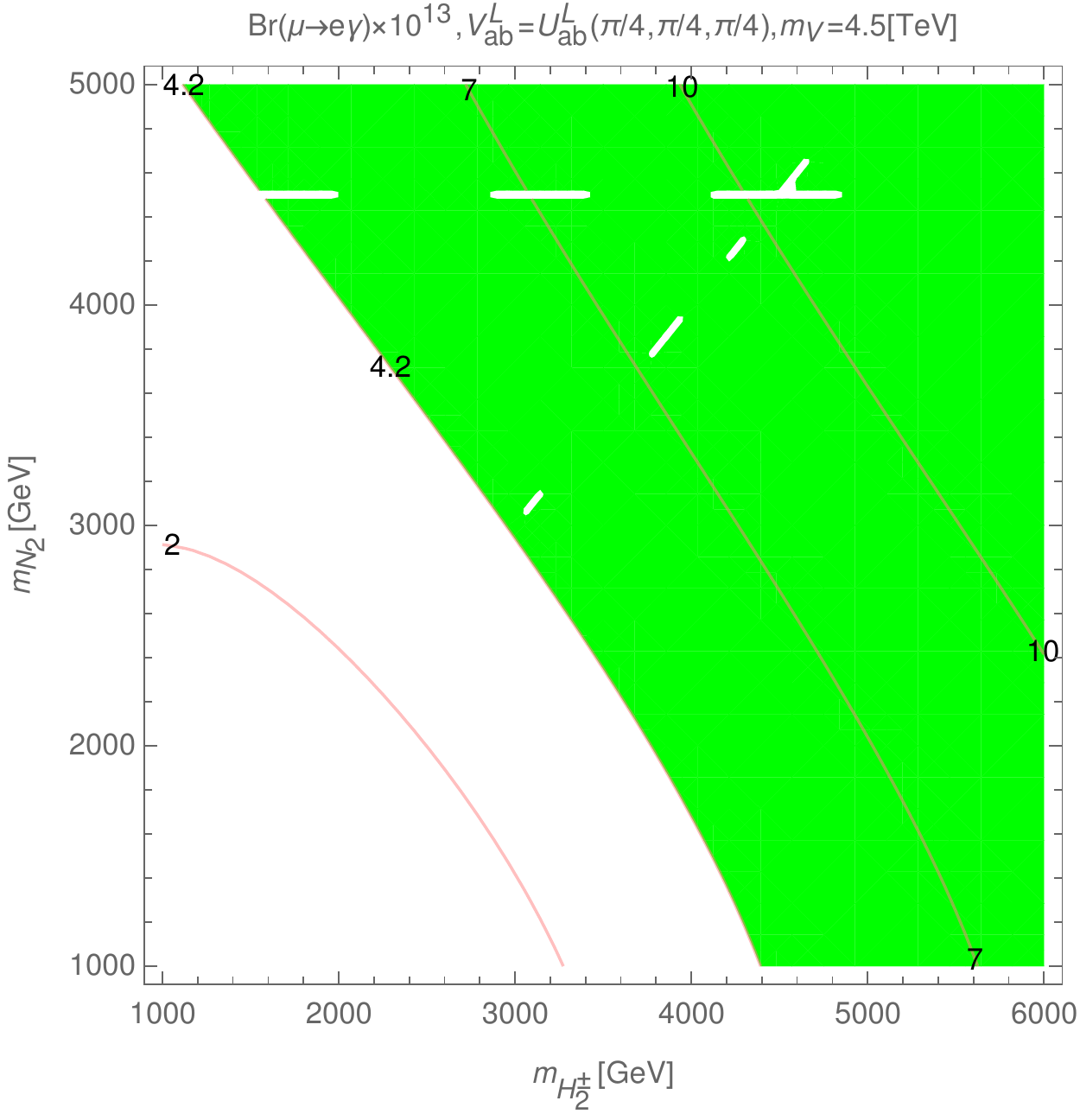}\\
		\includegraphics[width=7.0cm]{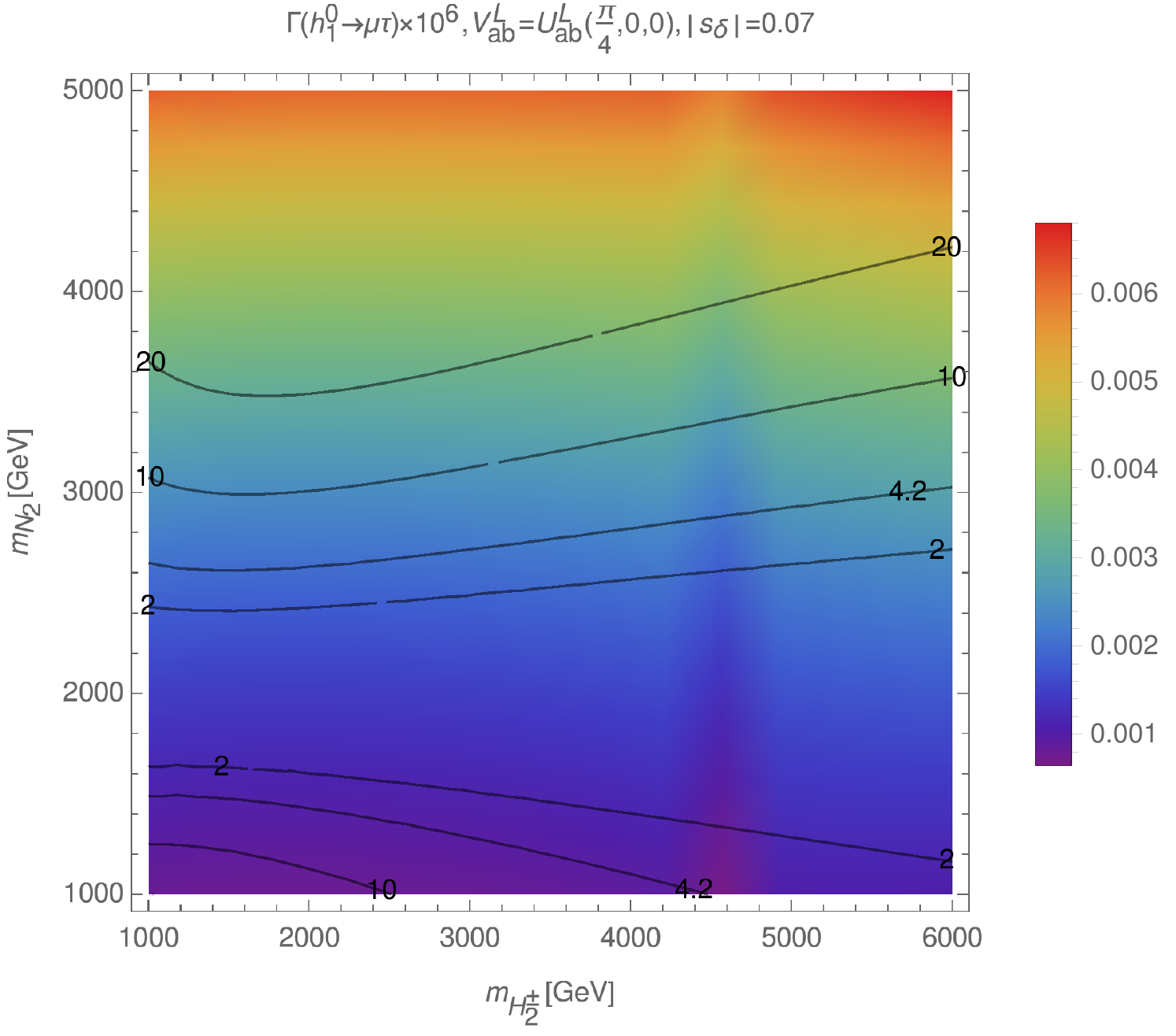}& \includegraphics[width=7.0cm]{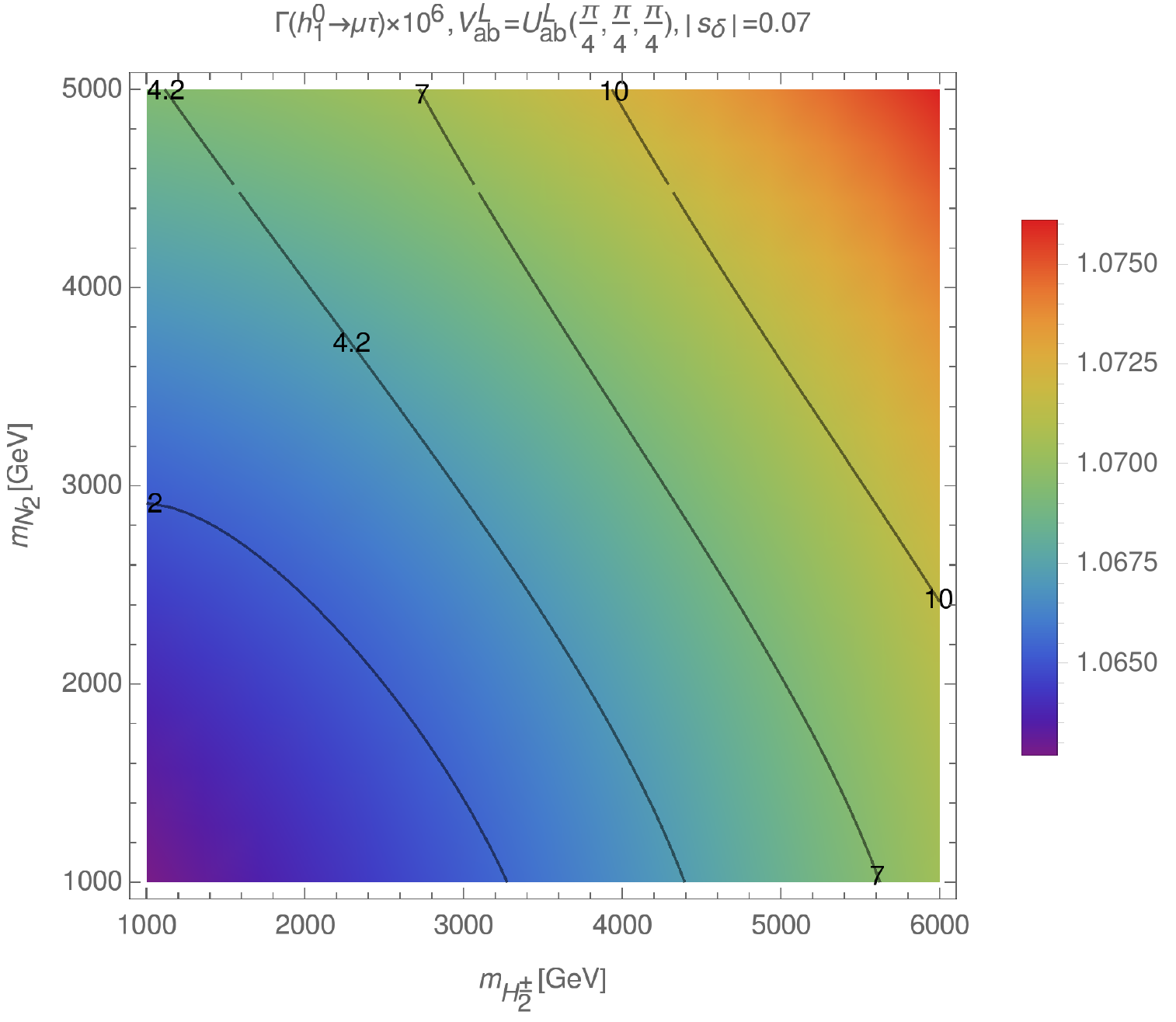}\\
	\end{tabular}%
	\caption{ Contour plots of $\mathrm{Br}(\mu \rightarrow e\gamma)$ ow and density plots of $\mathrm{\Gamma}(h^0_1 \rightarrow \mu\tau)$ (second row) as functions of $m_{H^\pm_2}$ and $m_{N_2}$ in the case of $V^L_{ab} = U^L_{ab}(\pi/4,0,0)$ (left) or in the case of  $V^L_{ab} = U^L_{ab}(\pi/4,\pi/4,\pi/4)$ (right). The black cuvers show the constant values of $\mathrm{Br}(\mu \rightarrow e\gamma)\times 10^{13}$.}
	\label{fig_raint3}
\end{figure}
%%%%%%%%%
The numerical results in the case of $V^L_{ab} = U^L_{ab}(\pi/4,0,0)$  and $V^L_{ab} = U^L_{ab}(\pi/4,\pi/4,\pi/4)$ are given in Fig.\ref{fig_raint3}. The signals of $\mathrm{\Gamma}(h^0_1 \rightarrow \mu\tau)$ in both cases are smaller than that shown in Fig.\ref{fig_raint1}. This statement is in agreement with what was previously published in Ref.\cite{Hung:2022kmv}.

%%%5%%%%
\begin{figure}[ht]
	\centering
	\begin{tabular}{cc}
	   	\includegraphics[width=7.0cm]{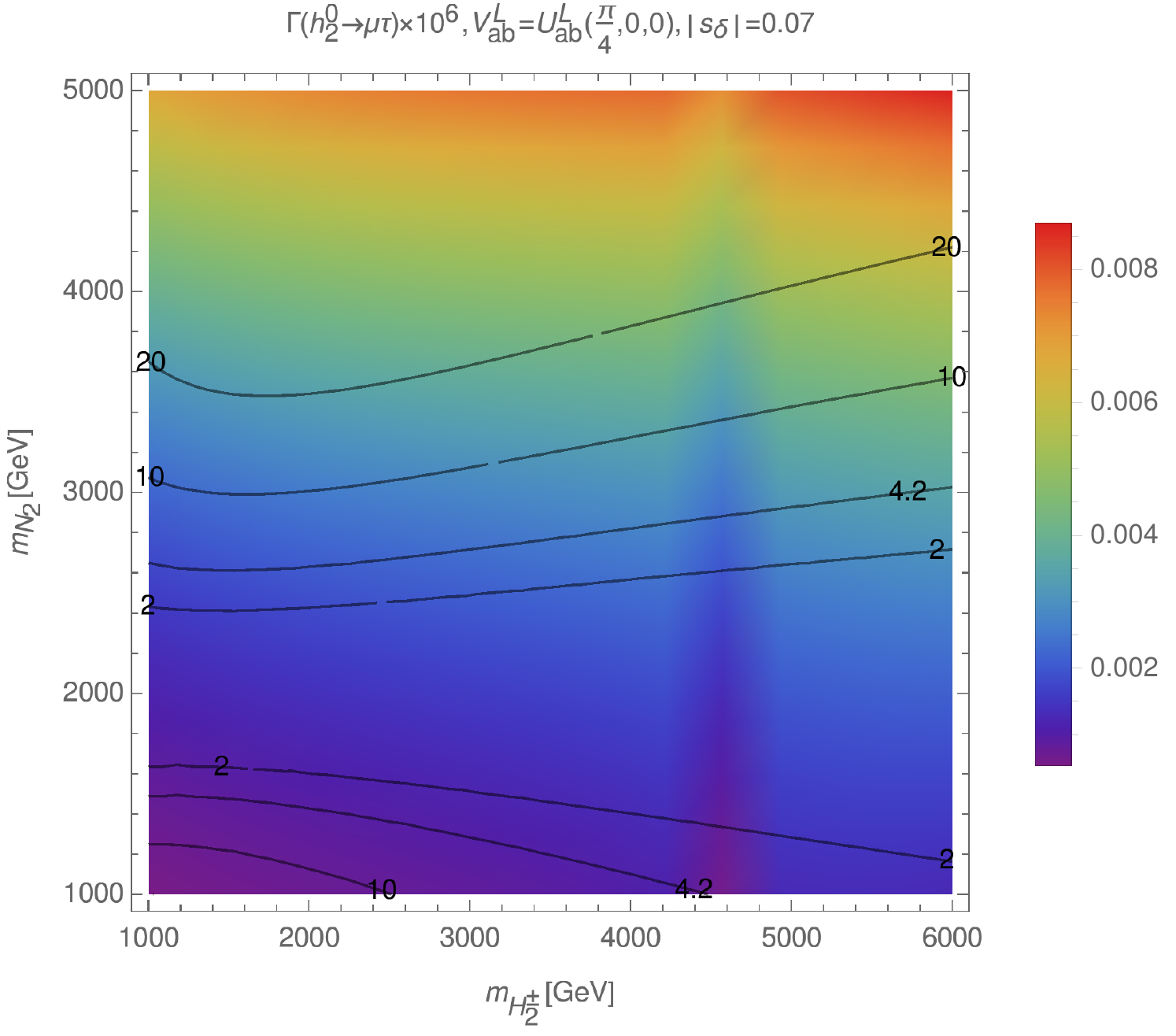}& \includegraphics[width=7.0cm]{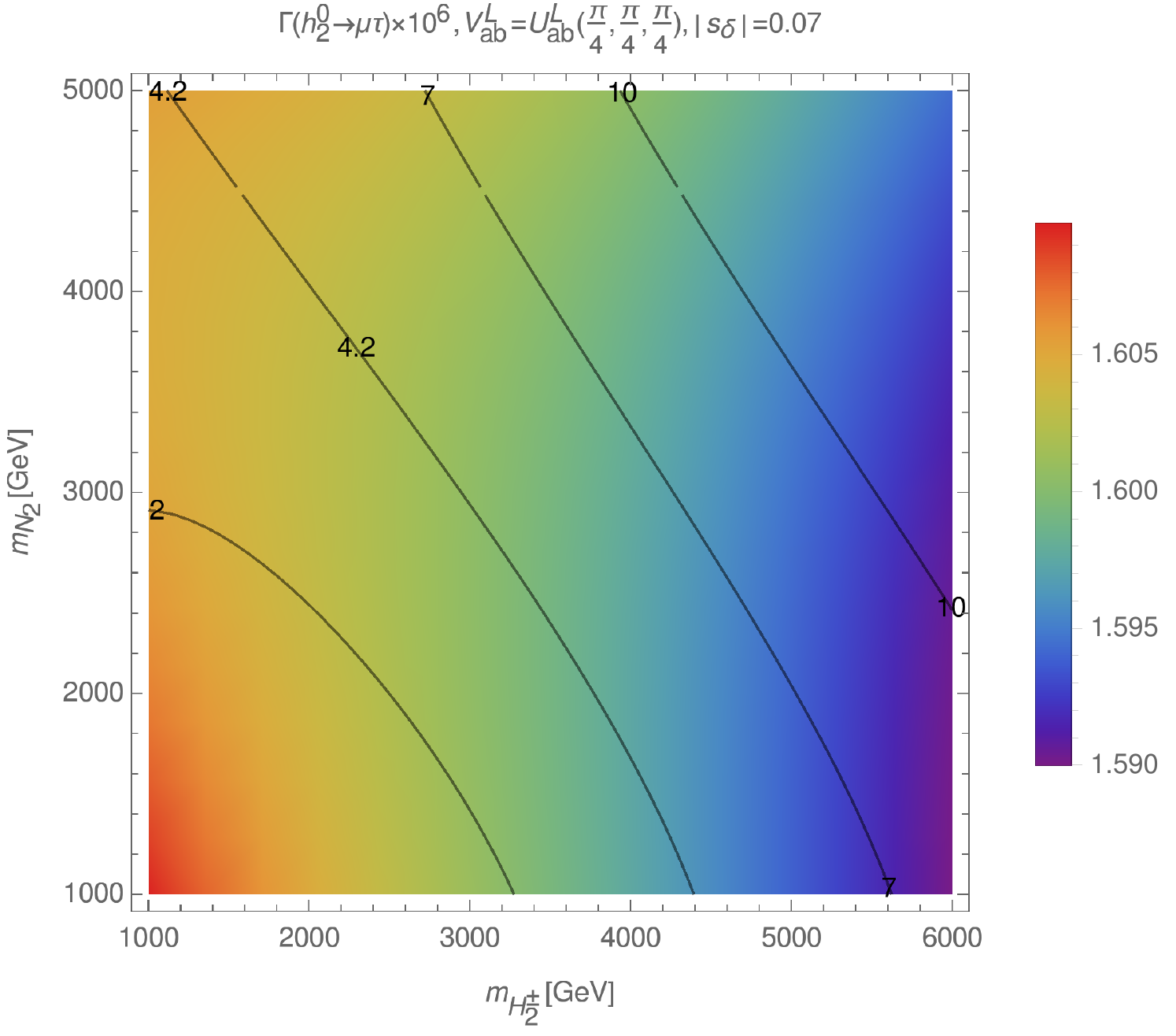}\\
		\includegraphics[width=7.0cm]{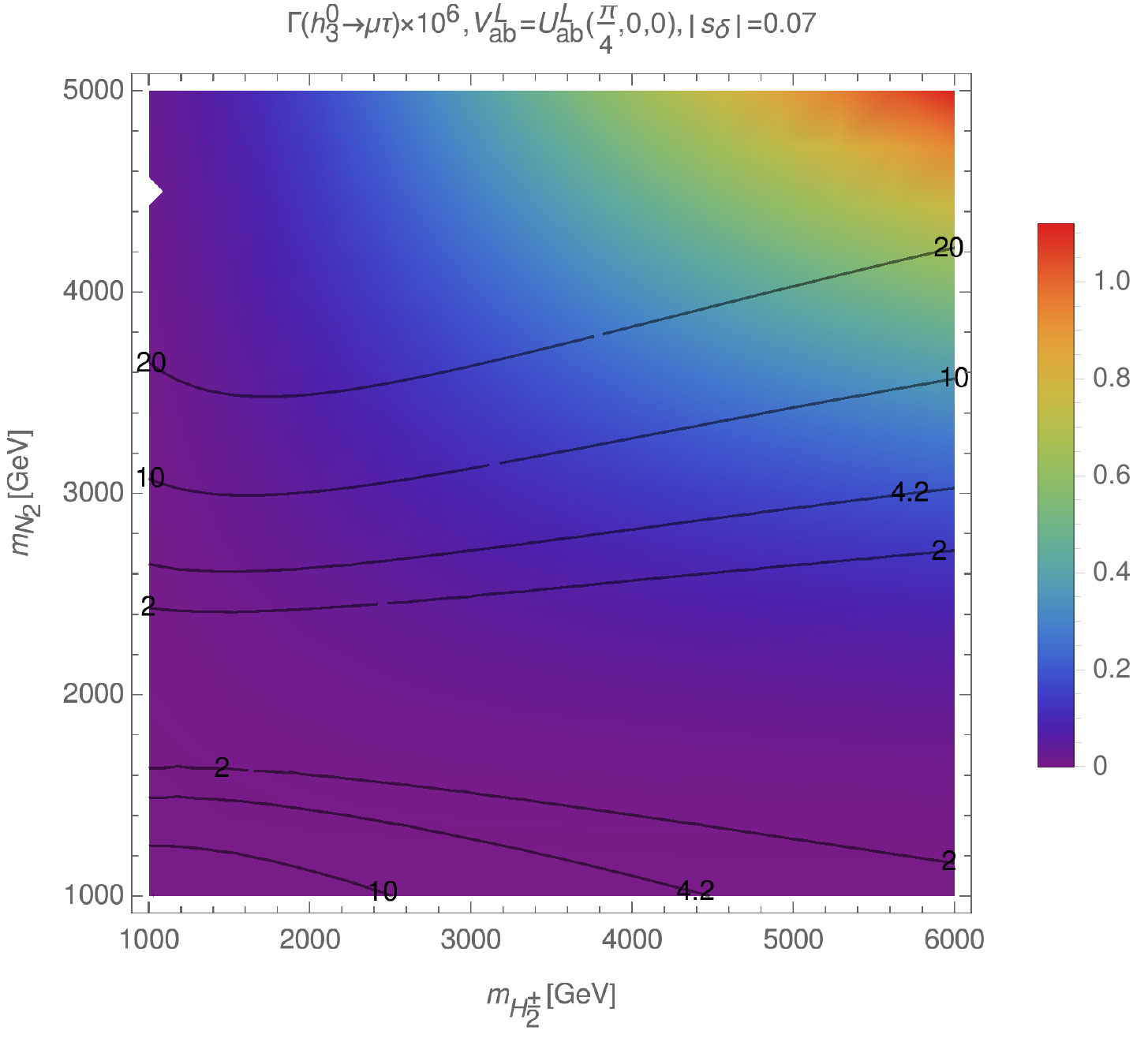}& \includegraphics[width=7.0cm]{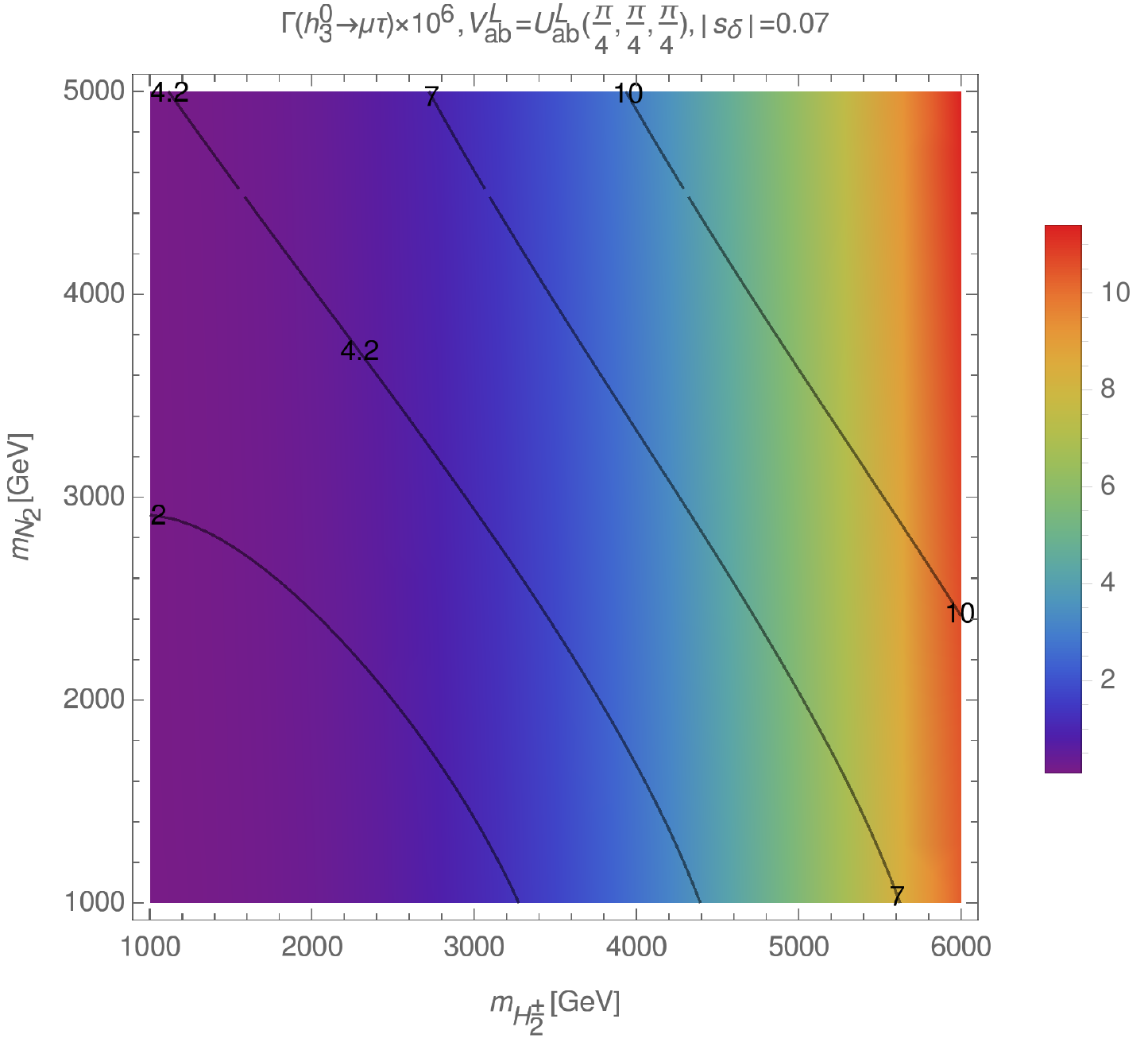}\\
	\end{tabular}%
	\caption{ Density plots of $\mathrm{\Gamma}(h^0_2 \rightarrow \mu\tau)$ (first row) $\mathrm{\Gamma}(h^0_3 \rightarrow \mu\tau)$ (second row) as functions of $m_{H^\pm_2}$ and $m_{N_2}$ in the case of $V^L_{ab} = U^L_{ab}(\pi/4,0,0)$ (left) or in the case of  $V^L_{ab} = U^L_{ab}(\pi/4,\pi/4,\pi/4)$ (right). The black cuvers show the constant values of $\mathrm{Br}(\mu \rightarrow e\gamma)\times 10^{13}$.}
	\label{fig_raint2}
\end{figure}
%%%%%%%%%
The signal of $\mathrm{\Gamma}(h^0_1 \rightarrow \mu\tau)$ depends very strongly on the form of the mixing matrix of neutral leptons. This has been confirmed as Ref.\cite{Hung:2022kmv}. In a similar way, we also show the dependence of $\mathrm{\Gamma}(h^0_2 \rightarrow \mu\tau)$ and $\mathrm{\Gamma}(h^0_3 \rightarrow \mu\tau)$ on the form of $V^L_{ab}$ as shown in Fig.\ref{fig_raint2}.\\
In the two cases $V^L_{ab} = U^L_{ab}(\pi/4,\pi/4,\pi/4)$ and $V^L_{ab} = U^L_{ab}(\pi/4,0,0)$, $\mathrm{\Gamma}(h^0_2 \rightarrow \mu\tau)$ and $\mathrm{\Gamma}(h^0_3 \rightarrow \mu\tau)$ are investigated in the parameter space regions bounded by $\mathrm{Br}(\mu \rightarrow e\gamma) < 4.2 \times 10^{13}$. The obtained results show that the signals of $\mathrm{\Gamma}(\mathrm{H} \rightarrow \mu\tau)$ can reach small values in the case of $V^L_{ab} = U^L_{ab}(\pi/4,0,0)$ and larger in the case of  $V^L_{ab} = U^L_{ab}(\pi/4,\pi/4,-\pi/4)$. This is completely consistent with previously mentioned publications such as Refs.\cite{Hung:2022kmv,Hue:2015fbb}. Based on Refs.~\cite{Patrignani:2016xqp,Zyla:2020zbs,CMS:2022ahq,Denner:2011mq}, if $\Gamma^\mathrm{total}_{\mathrm{H}}\simeq 4.1\times 10^{-3}~\mathrm{GeV}$ then $m_{h^0_1}=125.09\,[\mathrm{GeV}]$, we also derive from the results in Fig.~\ref{fig_raint3} and Fig.~\ref{fig_raint2} to predict $150~[GeV]\leq m_{h^0_2} \leq 190~[GeV]$ and $470~[GeV]\leq m_{h^0_3} \leq 3800~[GeV]$. These observations are expected to be confirmed experimentally soon.
%%%
\section{Conclusions}
\label{conclusion}
The 3-3-1 model with neutral leptons was put into the constraint of Higgs potential such as Eq.(\ref{eq_alignH0}) to avoid the tree level contributions of SM-like Higgs boson to the  flavor  changing neutral currents (FCNC) in the quark sector generated four CP-even Higgs bosons. The two lighter neutral Higgs are identified with their respective ones in 2HDM, an other with a larger mass that is not coupling with similar particles contained in the SM, only interacts with exotic particles and the remaining particle is assumed no having LFV couplings.\\
The contributions at one-loop order to $\mathrm{\Gamma}(\mathrm{H} \rightarrow \mu\tau)$  consists of two parts, that of active neutrinos and neutral leptons. However, the contribution of active neutrinos is so small compared to the corresponding part of neutral leptons that they can be ignored in some numerical investigation results.\\
The partial width ($\mathrm{\Gamma}(\mathrm{H} \rightarrow \mu\tau)$ ) depends strongly on the masses of charged Higgs boson ($m_{H^\pm_2}$), neutral leptons ($m_{N_2}$) as well as the mixing matrix form of the neutral leptons ($V^L_{ab}$).\\
$\mathrm{\Gamma}(\mathrm{H} \rightarrow \mu\tau)$ are investigated numerically in the regions of parameter space satisfying the experimental limit of the cLFV decays ($l_i \rightarrow l_j\gamma$). In these parameter space regions, the signal of the SM-like Higgs boson is always within the current experimental limit, $\mathrm{\Gamma}(h^0_1 \rightarrow \mu\tau) \leq 4.1 \times 10^{-6}$. The results also show that the signal of $\mathrm{\Gamma}(\mathrm{H} \rightarrow \mu\tau)$ can take small values in the case of $V^L_{ab} = U^L_{ab}(\pi/4,0,0)$ and larger in the cases of  $V^L_{ab} = U^L_{ab}(\pi/4,\pi/4,\pi/4)$ and $V^L_{ab} = U^L_{ab}(\pi/4,\pi/4,-\pi/4)$.\\
In addition, $\mathrm{\Gamma}(h^0_2 \rightarrow \mu\tau)$ is always very close to the corresponding ones of SM-like Higgs boson, on the contrary, $\mathrm{\Gamma}(h^0_3 \rightarrow \mu\tau)$ is very different from the previous two. It can be larger in domains $m_{H^\pm_2}>5[\mathrm{TeV}]$ or $m_{N_2}<20[\mathrm{TeV}]$ and smaller in domains $m_{H^\pm_2}<5[\mathrm{TeV}]$ or $m_{N_2}>20[\mathrm{TeV}]$. We also give predictions regarding the masses of the two heavier CP-even Higgs bosons in the 331NL model, $150~[GeV]\leq m_{h^0_2} \leq 190~[GeV]$ and $470~[GeV]\leq m_{h^0_3} \leq 3800~[GeV]$.  These signals of this model  are expected to be found in the near future when the accelerators operate at a larger energy scale.

\section*{Acknowledgments}
This research is funded by the Foundation for Science and Technology Development, Hanoi Pedagogical University 2 via grant number HPU2.2022-UT-04.

\appendix
\section{Master integrals.}
\label{appen_PV}
To calculate the contributions at one-loop order of the Feynman diagrams in Fig.\ref{fig_hmt331}, we use the Passarino-Veltman (PV) functions as mentioned in Ref.\cite{Passarino:1978jh}. By introducing the notations $D_0=k^2-M_0^2+i\delta$, $D_1=(k-p_1)^2-M_{1}^2+i\delta$ and $D_2=(k+p_2)^2-M_2^2+i\delta$, where $\delta$ is  infinitesimally a  positive real quantity, we have:
\bea
%----
A_{0}(M_n)
&\equiv &\frac{\left(2\pi\mu\right)^{4-D}}{i\pi^2}\int \frac{d^D k}{D_n}, \hs
B^{(1)}_0 \equiv\frac{\left(2\pi\mu\right)^{4-D}}{i\pi^2}\int \frac{d^D k}{D_0D_1},\crn
B^{(2)}_0 &\equiv &\frac{\left(2\pi\mu\right)^{4-D}}{i\pi^2}\int \frac{d^D k}{D_0D_2}, \hs
B^{(12)}_0 \equiv \frac{\left(2\pi\mu\right)^{4-D}}{i\pi^2}\int \frac{d^D k}{D_1D_2},\crn
C_0&\equiv&  C_{0}(M_0,M_1,M_2) =\frac{1}{i\pi^2}\int \frac{d^4 k}{D_0D_1D_2},
\label{scalrInte}\eea
where $n=1,2$, $D=4-2\epsilon \leq 4$ is the dimension of the integral, while $~M_0,~M_1,~M_2$ stand for the masses of virtual particles in the loops. We also assume  $p^2_1=m^2_{1},~p^2_2=m^2_{2}$ for external fermions. The tensor integrals are
\bea
%--
A^{\mu}(p_n;M_n)
&=&\frac{\left(2\pi\mu\right)^{4-D}}{i\pi^2}\int \frac{d^D k\times k^{\mu}}{D_n}=A_0(M_n)p_n^{\mu},\crn
B^{\mu}(p_n;M_0,M_n)&=& \frac{\left(2\pi\mu\right)^{4-D}}{i\pi^2}\int \frac{d^D k\times
	k^{\mu}}{D_0D_n}\equiv B^{(n)}_1p^{\mu}_n,\crn
%--
B^{\mu}(p_1,p_2;M_1,M_2)&=& \frac{\left(2\pi\mu\right)^{4-D}}{i\pi^2}\int \frac{d^D k\times
	k^{\mu}}{D_1D_2}\equiv B^{(12)}_1p^{\mu}_1+B^{(12)}_2p^{\mu}_2,\crn
%--
C^{\mu}(M_0,M_1,M_2)&=&\frac{1}{i\pi^2}\int \frac{d^4 k\times k^{\mu}}{D_0D_1D_2}\equiv  C_1 p_1^{\mu}+C_2 p_2^{\mu},\crn
C^{\mu \nu}(M_0,M_1,M_2)&=&\frac{1}{i\pi^2}\int \frac{d^4 k\times k^{\mu}k^{\nu}}{D_0D_1D_2}\equiv  C_{00}g^{\mu \nu}+C_{11} p_1^{\mu}p_1^{\nu}+C_{12} p_1^{\mu}p_2^{\nu}+C_{21} p_2^{\mu}p_1^{\nu}+C_{22} p_2^{\mu}p_2^{\nu},\crn
\label{oneloopin1}\eea
where $A_0$, $B^{(n)}_{0,1}$, $B^{(12)}_{n}$ and $C_{0,n}, C_{mn}$   are PV functions.  It is well-known that $C_{0,n}, C_{mn}$ are finite while the remains are divergent. We denote
\be \Delta_{\epsilon}\equiv \frac{1}{\epsilon}+\ln4\pi-\gamma_E, \label{divt}\ee with $\gamma_E$ is the  Euler constant.  

Using the technique as mentioned in Ref.\cite{Hue:2017lak}, we can show the divergent parts of the above PV functions as
\bea  \mathrm{Div}[A_0(M_n)]&=& M_n^2 \Delta_{\epsilon}, \hs  \mathrm{Div}[B^{(n)}_0]= \mathrm{Div}[B^{(12)}_0]= \Delta_{\epsilon}, \crn
\mathrm{Div}[B^{(1)}_1]&=&\mathrm{Div}[B^{(12)}_1] = \frac{1}{2}\Delta_{\epsilon},  \hs  \mathrm{Div}[B^{(2)}_1] = \mathrm{Div}[B^{(12)}_2]= -\frac{1}{2} \Delta_{\epsilon}.  \label{divs1}\eea
Apart from the divergent parts, the rest of these functions are finite.

Thus, the above PV functions can be written in form:
\be  A_0(M)= M^2\Delta_{\epsilon}+a_0(M),\,\, B^{(n)}_{0,1}= \mathrm{Div}[B^{(n)}_{0,1}]+ b^{(n)}_{0,1}, \,\,  B^{(12)}_{0,1,2}= \mathrm{Div}[B^{(12)}_{0,1,2}]+ b^{(12)}_{0,1,2}, \label{B01i}\ee
where $a_0(M), \,\,b^{(n)}_{0,1}, \,\, b^{(12)}_{0,1,2} $ are finite parts and have a specific form defined as Ref.\cite{Thuc:2016qva} for $\mathrm{H} \rightarrow \mu \tau$ decay.
\section{Analytic formulas at one-loop order  to $\mathrm{H} \rightarrow l_il_j$.}
\label{appen_loops1}
The one-loop factors of the diagrams in Fig.(\ref{fig_hmt331}) are given in this appendix. We used the same calculation techniques as shown in \cite{Thuc:2016qva,Phan:2016ouz}. We denote $m_{l_i} \equiv  m_1$ and $ m_{l_j} \equiv m_2$.
%-- diagram 1)
{\small\bea  \mathcal{A}^{(1)}_L(m_F,m_V)
	&=&m_V m_1\left\{ \frac{1}{2m_V^4}\left[m_F^2(B^{(1)}_1-B^{(1)}_0-B^{(2)}_0)\right.\right.\crn
	\hs &-&\left.\left. m_2^2B^{(2)}_1 + \left(2m_V^2+m^2_{H}\right)m_F^2\left(C_0-C_1\right)\right]\right.\crn
	&&\left.-\left(2+\frac{m_1^2-m_2^2}{m_V^2}\right) C_1 +
	\left(\frac{m_1^2-m^2_{H}}{m_V^2}+ \frac{m_2^2 m^2_{H}}{2m_V^4}\right)C_2\right\}, \label{EfvvL} \\
	%---ER
	\mathcal{A}^{(1)}_R(m_F,m_V)&=&m_V m_2\left\{\frac{1}{2 m_V^4}\left[-m_F^2\left(B^{(2)}_1+ B^{(1)}_0 + B^{(2)}_0 \right) \right.\right.\crn
	&+& \left.\left.  m_1 ^2 B^{(1)}_1  +   (2m_V^2+m^2_{H}) m_F^2(C_0+C_2)\right] \right.\crn
	&&\left.+\left(2+\frac{-m_1^2+m_2^2}{m_V^2}\right)C_2-\left( \frac{m_2^2-m^2_{H}}{m_V^2}+ \frac{m_1^2 m^2_{H}}{m_V^4}\right)C_1\right\},  \label{EfvvR}
	\eea
	%---- Diagram 2)
	\bea
	&& \mathcal{A}^{(2)}_L(a_1,a_2,v_1,v_2,m_F,m_V,m_{H^\pm})\crn&=&
	m_1\left\{-\fr{a_2}{v_2} \fr{m_F^2}{m_V^2}\left(B^{(1)}_1-B^{(1)}_0\right)   + \fr{a_1}{v_1}m_2^2\left[2 C_1-\left(1+ \fr{m^2_{H^\pm}-m^2_{H}}{m_V^2}\right) C_2\right]\right.\crn
	&&\left.+\fr{a_2}{v_2}m_F^2\left[C_0+C_1+\fr{m^2_{H^\pm}-m^2_{H}}{m_V^2}\left(C_0-C_1\right) \right]\right\}, \label{EfvhL} \\
	%%%%%%%%%%%%
	&& \mathcal{A}^{(2)}_R(a_1,a_2,v_1,v_2,m_F,m_V,m_{H^\pm})\crn &=&m_2\left\{\fr{a_1}{v_1}\left[\fr{m_1^2B^{(1)}_1-m_F^2B^{(1)}_0}{m_V^2} +\left(\frac{}{}m_F^2C_0-m_1^2C_1+2 m_2^2C_2\right.\right.\right.\crn
	&&\left.\left.+2(m^2_{H}-m_2^2)C_1-  \fr{m^2_{H^\pm}-m^2_{H}}{m_V^2}\left(m^2_FC_0-m_1^2C_1\right)\right)\right]\crn &&+\left.\fr{a_2}{v_2} m_F^2\left(-2C_0-C_2+\fr{m^2_{H^\pm}-m^2_{H}}{m_V^2}C_2 \right) \right\},   \label{EfvhR}
	\eea
	%---- Diagram 3)
	\bea
	&& \mathcal{A}^{(3)}_L(a_1,a_2,v_1,v_2,m_F,m_{H^\pm},m_V)\crn&=& m_1\left\{\fr{a_1}{v_1}\left[\fr{-m_2^2B^{(2)}_1-m_F^2B^{(2)}_0}{m_V^2} +\left(\frac{}{}m_F^2C_0-2m_1^2C_1+ m_2^2C_2\right.\right.\right.\crn
	&&\left.\left.-2(m^2_{H}-m_1^2)C_2-  \fr{m^2_{H^\pm}-m^2_{H}}{m_V^2}\left(m^2_FC_0+m_2^2C_2\right)\right)\right] \crn &&\left.+\fr{a_2}{v_2} m_F^2\left(-2C_0+C_1-\fr{m^2_{H^\pm}-m^2_{H}}{m_V^2}C_1 \right)\right\}, \label{EfhvL} \\
	%%%%%%%%%%%%
	&& \mathcal{A}^{(3)}_R(a_1,a_2,v_1,v_2,m_F,m_{H^\pm},m_V)\crn&=& m_2 \left\{\fr{a_2}{v_2} \fr{m_F^2}{m_V^2}\left(B^{(2)}_1+B^{(2)}_0\right)
	+ \fr{a_1}{v_1}m_1^2\left[-2 C_2+\left(1+ \fr{m^2_{H^\pm}-m^2_{H}}{m_V^2}\right) C_1\right]\right.\crn
	&&\left.+\fr{a_2}{v_2}m_F^2\left[C_0-C_2+\fr{m^2_{H^\pm}-m^2_{H}}{m_V^2}\left(C_0+C_2\right) \right]\right\}.   \label{EfhvR}
	\eea
		%---- Diagram 4)+5)
	\bea
	\mathcal{A}^{(4+5)}_L(m_F,m_V)&=& \fr{-m_1m_2^2}{m_V(m_1^2-m_2^2)}\left[\left(2+\frac{m_F^2}{m_V^2}\right) \left(B^{(1)}_1 +B^{(2)}_1 \right) \right. \crn&+&\left.\fr{m_1^2 B^{(1)}_1 +m_2^2 B^{(2)}_1}{m_V^2} - \fr{2m_F^2}{m_V^2}\left(B^{(1)}_0-B^{(2)}_0\right)\right],  \label{DfvL} \\
	\mathcal{A}^{(4+5)}_R(m_F,m_V)&=& \frac{m_1}{m_2}E^{FV}_L, \label{DfvR}\eea
		%---- Diagram 6)
	%----
	\bea
	\mathcal{A}^{(6)}_L(a_1,a_2,v_1,v_2,m_F,m_{H^\pm})&=&\frac{ m_1m^2_F }{v_2}\crn
	&\times& \left[\dfrac{a_1a_2}{v_1v_2}B^{(12)}_{0}
	+\fr{a_1^2}{v_1^2}m_2^2(2C_2+C_0)+\fr{a_2^2}{v_2^2}m_F^2(C_0-2C_1) \right.\crn
	&+& \left.\fr{a_1a_2}{v_1v_2} \left(\frac{}{}2m_2^2C_2-(m_1^2+m_2^2)C_1+(m_F^2+m^2_{H^\pm}+m_2^2)C_0\right)\right],\crn  \label{EhffL} \\
	\mathcal{A}^{(6)}_R(a_1,a_2,v_1,v_2,m_F,m_{H^\pm})&=& \frac{m_2 m^2_F}{v_2}\crn
	&\times&\left[ \dfrac{a_1a_2}{v_1v_2}B^{(12)}_{0}+ \dfrac{a_1^2}{v_1^2}m_1^2(C_0-2C_1)+\fr{a_2^2}{v_2^2}m_F^2(C_0+2C_2)\right.\crn
	&+&\left. \fr{a_1a_2}{v_1v_2}\left(\frac{}{}-2m_1^2C_1+(m_1^2+m_2^2)C_2+(m_F^2+m^2_{H^\pm}+m_1^2)C_0 \right)\right], \crn\label{EhffR} \eea
	%---- Diagram 7)
	\bea
	\mathcal{A}^{(7)}_L(a_1,a_2,v_1,v_2,m_F,m_{H^\pm})&=&  m_1v_2\left[ \fr{a_1a_2}{v_1v_2}m_F^2C_0-\fr{a^2_1}{v^2_1}m_2^2C_2+\fr{a^2_2}{v^2_2}m_F^2C_1\right] , \crn \label{EfhhL} \\
	\mathcal{A}^{(7)}_R(a_1,a_2,v_1,v_2,m_F,m_{H^\pm})&=& m_2v_2\left[ \fr{a_1a_2}{v_1v_2}m_F^2C_0+\fr{a^2_1}{v^2_1}m_1^2C_1-\fr{a^2_2}{v^2_2}m_F^2C_2 \right],\crn \label{EfhhR} \eea
	%---- Diagram 8)
	\bea
	\mathcal{A}^{(8)}_L(m_V,m_F)&=&\frac{m_1m^2_F}{m_V}\crn
	&\times&\left[\fr{1}{m_V^2}\left(B^{(12)}_{0}
	+B^{(1)}_1 -(m_1^2+m_2^2-2m_F^2)C_1\right)-C_0+4C_1\right],\crn  \label{EvffL} \\
	\mathcal{A}^{(8)}_R(m_V,m_F)&=&
	\frac{m_2 m^2_F}{m_V} \crn
	&\times&\left[ \fr{1}{m_V^2}\left(B^{(12)}_{0} -B^{(2)}_1 +(m_1^2+m_2^2-2m_F^2)C_2\right)-C_0-4C_2 \right],\crn \label{EvffR}\eea
	%%

%%%%%%%%%%%% diagram 9+10%%%%%%%%%%%%
	\bea
	\mathcal{A}^{(9+10)}_L(a_1,a_2,v_1,v_2,m_F,m_{H^\pm})&=& \fr{m_1}{v_1(m_1^2-m_2^2)}\left[m^2_2 \left(m^2_1\fr{a_1^2}{v_1^2}+m^2_F\fr{a^2_2}{v^2_2} \right)
	\left(B_1^{(1)}+B_1^{(2)}\right) \right.\crn
	&&\left.  \hspace{1.8 cm}+m^2_F\fr{a_1a_2}{v_1v_2}\left(2m^2_2B_0^{(1)}-(m^2_1 +m^2_2)B_0^{(2)}\right) \right], \label{DfhL} \\
	\mathcal{A}^{(9+10)}_R(a_1,a_2,v_1,v_2,m_F,m_{H^\pm})&=&  \fr{m_2}{v_1(m_1^2-m_2^2)}\left[ m^2_1 \left(m^2_2\fr{a_1^2}{v_1^2}+m^2_F\fr{a^2_2}{v^2_2} \right)\left(B_1^{(1)}+B_1^{(2)}\right)\right.\crn
	&&\left.  \hspace{1.8 cm}+m^2_F\fr{a_1a_2}{v_1v_2}\left(-2m^2_1B_0^{(2)}+(m^2_1 +m^2_2)B_0^{(1)}\right)\right]. \label{DfhR} \eea}
Obviously, with the 7th diagram of Fig.(\ref{fig_hmt331}) we always get that $\mathcal{A}^{(7)}_{L,R}(a_1,a_2,v_1,v_2,m_F,m_H)$ are finite because $C_i$ ($i=0,1,2$) do not contain divergent functions.
%%%%%%%%
\section{\label{CaDV} The analytic formulas and divergent cancellation in amplitudes}
\label{appen_loops2}
Similar to the case of $h^0_1$ in Sec.\ref{Analytic}, here we give analytic formulas  and divergence canceling in the amplitudes of the decay channels to $\mu \tau$ of $h^0_2$ and $h^0_3$.

The contribution to the decay of $h^0_2 \rightarrow \mu\tau$ consists of two parts: the first part is the contribution of ordinary neutrinos and the second part is the contribution of neutral leptons. 

The first part has the appearance of all the diagrams in Fig.\ref{fig_hmt331}, the analytic expression is

%--For active neutrino
\bea  \Delta^{(ij)\nu-h^0_2}_{L,R} &=&  \sum_{a}U_{ia} U_{ja}^{*} \frac{1}{64\pi^2}\left[2g^3(c_{\alpha}c_{12}+s_\alpha s_{12})\times \mathcal{A}^{(1)}_{L,R}(m_{\nu_a},m_W)\right. \crn
&&-g^2(c_{\alpha}s_{12}-s_\alpha c_{12})\times \mathcal{A}^{(2)}_{L,R} (s_{12},c_{12},v_1,v_2,m_{\nu_a},m_W,m_{H^{\pm}_1}) \crn
&&-g^2(c_{\alpha}s_{12}-s_\alpha c_{12})\times \mathcal{A}^{(3)}_{L,R} (s_{12},c_{12},v_1,v_2,m_{\nu_a},m_W,m_{H^{\pm}_1}) \crn
&&-g^3c_{\alpha}\times \mathcal{A}^{(4+5)}_{L,R}(m_{\nu_a},m_W)\crn
&&-2s_{\alpha}\times \mathcal{A}^{(6)}_{L,R}(s_{12},c_{12},v_1,v_2,m_{\nu_a},m_{H^{\pm}_1})\crn
&&-2 \lambda_{h^0_2H_1H_1}\times \mathcal{A}^{(7)}_{R} (s_{12},c_{12},v_1,v_2,m_{\nu_a},m_{H^{\pm}_1})\crn
&&-(g^3s_{\alpha}/2)\times \mathcal{A}^{(8)}_{L,R}(m_W,m_{\nu_a})\crn
&&-\left.  c_{\alpha} \times \mathcal{A}^{(9+10)}_{L,R} (s_{12},c_{12},v_1,v_2,m_{\nu_a},m_{H^{\pm}_1})\right].  \label{nudeltaL2}\eea
%%%%%%%
and the second part contains only diagrams (1), (2), (3), (4), (5), (7), (9), (10) in Fig.\ref{fig_hmt331}, its analytic representation is
%%%%%%%
\bea  \Delta^{(ij)N-h^0_2}_{L,R} &=& \sum_{a}V_{ia}^LV_{ja}^{L*}  \frac{1}{64 \pi^2}
%---exotic lepton
\left[(-2g^3c_\al s_{13}) \times \mathcal{A}^{(1)}_{L,R}(m_{N_a},m_V) \right.\crn
&&-g^2c_{\alpha}c_{13}\times \mathcal{A}^{(2)}_{L,R} (c_{13},s_{13},v_1,v_3,m_{N_a},m_V,m_{H^{\pm}_2}) \crn
&&-g^2c_{\alpha}c_{13}\times \mathcal{A}^{(3)}_{L,R} (c_{13},s_{13},v_1,v_3,m_{N_a},m_V,m_{H^{\pm}_2}) \crn
&&-g^3c_{\alpha}\times \mathcal{A}^{(4+5)}_{L,R}(m_{N_a},m_V)\crn
&&- 2\lambda_{h^0_2H_2H_2} \times \mathcal{A}^{(7)}_{L,R} (c_{13},s_{13},v_1,v_3,m_{N_a},m_{H^{\pm}_2})\crn
&&-\left. c_{\alpha}\times \mathcal{A}^{(9+10)}_{L,R} (c_{13},s_{13},v_1,v_3,m_{N_a},m_{H^{\pm}_2})\right].  \label{NdeltaL2}\eea
%%%%%%%%%%%%%%%%

Using the formulas in Appendix. \ref{appen_loops1} and the notations as given in Eq.(\ref{aijnulepton}), we show that the divergence terms in the contribution of the neutral leptons to the $h^0_2 \rightarrow \mu\tau$ decay are
%%%%%%%%%%%%%%
\bea \mathrm{Div}\left[\Delta_{(1)}^{(ij)-h^0_2}\right] &=& B\times (-3)c_{\alpha}s_{13}, \crn   \mathrm{Div}\left[\Delta_{(2+3)}^{(ij)-h^0_2}\right]&=&  B\times c_{\alpha}\left(3s_{13} -\frac{2}{s_{13}}\right), \crn
\mathrm{Div}\left[\Delta^{(ij)-h^0_2}_{(4)} \right]&=&\fr{1}{m_1^2-m_2^2}\left[m_2^2B_L+m_1^2B_R\right]\times \fr{-c_\al}{s_{13}},\crn
\mathrm{Div}\left[\Delta^{(ij)-h^0_2}_{(5)} \right]&=&\fr{1}{m_1^2-m_2^2}\left[m_2^2B_L+m_1^2B_R\right]\times \fr{c_\al}{s_{13}},\crn
\mathrm{Div}\left[\Delta_{(9+10)}^{(ij)-h^0_2}\right]&=& B\times  c_{\alpha}\times\frac{2}{s_{13}}, \label{canceldiv3}\eea
%----
where $B$, $B_L$ and $B_R$ are shown in Eq.(\ref{B_Na}). It is easy to see that the sum over all factors is zero. 

For contribution of the active  neutrinos, we use the notations in Eq.(\ref{aijNlepton}) to give the divergence terms.
\bea
\mathrm{Div}\left[\Delta^{(ij)}_{(1)} \right]&=&\mathcal{B}\times (-3)\left( c_\al c_{12}+s_\al s_{12}\right) ,\crn
\mathrm{Div}\left[\Delta^{(ij)}_{(2)} \right]&=&\mathcal{B}\times \frac{-3}{2}\left( c_\al s_{12}-s_\al c_{12}\right)\frac{s_{12}}{c_{12}},\crn
\mathrm{Div}\left[\Delta^{(ij)}_{(3)} \right]&=&\mathcal{B}\times \frac{3}{2}\left( c_\al s_{12}-s_\al c_{12}\right)\frac{c_{12}}{s_{12}},\crn
\mathrm{Div}\left[\Delta^{(ij)}_{(4)} \right]&=&\fr{1}{m_1^2-m_2^2}\left[m_2^2\mathcal{B}_L+m_1^2\mathcal{B}_R\right]\times (\frac{c_\al}{s_{12}}),\crn
\mathrm{Div}\left[\Delta^{(ij)}_{(5)} \right]&=&\fr{1}{m_1^2-m_2^2}\left[m_2^2\mathcal{B}_L+m_1^2\mathcal{B}_R\right]\times (\frac{-c_\al}{s_{12}}),\crn
\mathrm{Div}\left[\Delta^{(ij)}_{(6+8)}\right]&=&\mathcal{B}\times (\frac{-3s_\al}{2s_{12}}),\crn
\mathrm{Div}\left[\Delta^{(ij)}_{(9+10)} \right]&=&\mathcal{B}\times (\frac{3c_\al}{2c_{12}}), \label{canceldiv4} \eea
where $\mathcal{B}$, $\mathcal{B}_L$ and $\mathcal{B}_R$ are shown in Eq.(\ref{B_Na}). It is easy to see that the sum over all factors is zero. 

%%%%%%%h03%%%%%
From Tab.\ref{albga}, we see that $h^0_3$ is not coupled to active neutrinos and $W^\pm$-bosons, thus $h^0_3 \rightarrow \mu\tau$ only contain diagrams (1), (2), (3), (6), (7), (8), (9) in Fig.~\ref{fig_hmt331}. The analytic expression is:

%%%%%%%
\bea  \Delta^{(ij)N-h^0_3}_{L,R} &=& \sum_{a}V_{ia}^LV_{ja}^{L*}  \frac{1}{64 \pi^2}
%---exotic lepton
\left[2g^3c_{23} \times \mathcal{A}^{(1)}_{L,R}(m_{N_a},m_V) \right.\crn
&&+g^2s_{23}\times \mathcal{A}^{(2)}_{L,R} (c_{23},s_{23},v_2,v_3,m_{N_a},m_V,m_{H^{\pm}_2}) \crn
&&+ g^2s_{23}\times \mathcal{A}^{(3)}_{L,R} (c_{23},s_{23},v_2,v_3,m_{N_a},m_V,m_{H^{\pm}_2}) \crn
&&-2\times \mathcal{A}^{(6)}_{L,R}(m_{N_a},m_V)\crn
&&- 2\lambda_{h^0_3H_2H_2} \times \mathcal{A}^{(7)}_{L,R} (c_{23},s_{23},v_2,v_3,m_{N_a},m_{H^{\pm}_2})\crn
&&-\left. 2g^3\times \mathcal{A}^{(8)}_{L,R} (c_{23},s_{23},v_2,v_3,m_{N_a},m_{H^{\pm}_2})\right]  \label{NdeltaL3}\eea
%%%%%%%%%%%%%%%%

And we concentrate on the divergent parts which are presented in the amplitudes calculated above. With the notations of the divergences shown in the appendix \ref{appen_loops1}, all of  divergent parts are collected as follows,
\bea \mathrm{Div}\left[\Delta_{(1)}^{(ij)}\right] &=& B\times (-3)c_{23}, \crn   \mathrm{Div}\left[\Delta_{(2+3)}^{(ij)}\right]&=&  B\times \left(3c_{23} -\frac{1}{c_{23}}\right), \crn
\mathrm{Div}\left[\Delta^{(ij)}_{(6+8)} \right]&=&B\times \left(\frac{1}{c_{23}}\right), \label{canceldiv5}\eea
%----
where  $B$ is mentioned in Eq.(\ref{B_Na}). We see again that sum of all divergent terms is zero.
%%%%%%%%%%%%%%%%%%%

Although, $h^0_3$ does not coupling with active neutrinos, it combines with $H_1^\pm$ Tab.~\ref{albga}. Therefore, the only contribution of active neutrinos  to h03mt is the diagram (7) in Fig.~\ref{fig_hmt331}. The result is :
\bea  \Delta^{(ij)\nu-h^0_3}_{L,R} = - \sum_{a}U_{ia} U_{ja}^{*} \frac{\lambda_{h^0_3H_1H_1}}{32\pi^2}\left[
 \times \mathcal{A}^{(7)}_{L,R} (s_{12},c_{12},v_1,v_2,m_{\nu_a},m_{H^{\pm}_1})\right] 
\label{Deltnuh3},
\eea
%%%%%%%
This term is finite since $\mathcal{A}^{(7)}_{L,R} (s_{12},c_{12},v_1,v_2,m_{\nu_a},m_{H^{\pm}_1})$ contains no divergence.

%%%%%%%%%%%%
%-----
 %\bibliographystyle{h-physrev}
%\bibliography{mainHQ2}

\end{document}